\input harvmac

\def\la{{\lambda}}
\def\Si{{\Sigma}}
\def\si{{\sigma}}
\def\al{{\alpha}}
\def\x{\times}
\def\ox{\otimes}
\def\cC{{\cal C}}
\def\cO{{\cal O}}
\def\cE{{\cal E}}
\def\F{{\cal F}}
\def\G{{\cal G}}
\def\cV{{\cal V}}
\def\cL{{\cal L}}
\def\cM{{\cal M}}

\def\ua{{\uparrow}}
\def\ra{{\rightarrow}}
\def\da{{\downarrow}}
\def\lra{{\longrightarrow}}

\noblackbox


\Title{ {\vbox{ \rightline{\hbox{hep-th/0703210}}
\rightline{\hbox{LMU-ASC 19/07}} }}} {\vbox{ \hbox{\centerline{
 Extension Bundles and the Standard Model}}}}
\centerline{Bj\"orn Andreas\foot{supported by 
DFG-SFB 647/A1.}$^{2}$~and~Gottfried Curio$^3$}
\bigskip
\centerline{$^2$ \it Institut f\"ur Mathematik und Informatik, 
Freie Universit\"at Berlin}
\centerline{\it Arnimallee 14, 14195 Berlin, Germany}
\smallskip
\centerline{$^3$ \it Arnold-Sommerfeld-Center for Theoretical Physics}
\centerline{\it Department f\"ur Physik, 
Ludwig-Maximilians-Universit\"at M\"unchen}
\centerline{\it Theresienstr. 37, 80333 M\"unchen, Germany}
\bigskip

\bigskip\bigskip
\noindent
We construct a new class of stable vector bundles suitable 
for heterotic string compac\-tifications. 
Using these we describe a novel way to derive the fermionic matter content 
of the Standard Model from the heterotic string. More precisely, we can get either 
the Standard Model gauge group $G_{SM}$ times an additional $U(1)$, or just $G_{SM}$ but 
with additional exotic matter. 
For this we compactify on an elliptically fibered Calabi-Yau threefold $X$
with two sections, the $B$-fibration, 
a variant of the ordinary Weierstrass fibration, 
which allows $X$ to carry a free involution. 
We construct rank five vector bundles, 
invariant under this involution, 
such that turning on a Wilson line 
we obtain the Standard Model gauge group 
and find various three generation models. 
This rank five bundle is derived from a stable rank four bundle that arises 
as an extension of bundles pulled-back from the base and
twisted by suitable line bundles. 
We also give an account of various previous results and put the present 
construction into perspective.

\Date{March 2007}

\lref\Artamk{I.V. Artamkin,{\it Deforming torsion-free sheaves 
on an algebraic surface},
Math. USSR Izv. {\bf 36} (1991), 449.}

\lref\Li{W.P. Li, Z.B. Qin, {\it Stable vector bundles 
on algebraic surfaces}, Trans. AMS 345 (1994), 833.}

\lref\Hart{R. Hartshorne, {\it Algebraic Geometry}, Springer 1977.}

\lref\BfibreI{B. Andreas, G. Curio and A. Klemm,
{\it Towards the Standard Model spectrum from elliptic Calabi-Yau},
hep-th/9903052, Int.J.Mod.Phys. {\bf A19} (2004) 1987.}

\lref\BfibreII{G. Curio, {\it Standard Model bundles of the heterotic string},
hep-th/0412182, Int.J.Mod.Phys. {\bf A21} (2006) 1261.}

\lref\BfibreIII{B. Andreas and G. Curio, 
{\it Invariant Bundles on $B$-fibered Calabi-Yau spaces 
and the Standard Model}, hep-th/0602247.}

\lref\extpaper{B. Andreas and G. Curio,
{\it Stable bundle extensions on elliptic Calabi-Yau threefolds}, 
math.AG/0611762.}

\lref\physpaper{B. Andreas and G. Curio, 
{\it Heterotic Models without Fivebranes}, hep-th/0611309.}

\lref\A{B. Andreas, 
{\it On vector bundles and chiral matter in N=1 heterotic compactifications},
JHEP {\bf 9901} (1999) 011, hep-th/9802202.}

\lref\Distler{J.Distler and B.R. Greene, {\it Aspects of (2,0) String
Compactifications}, Nucl. Phys {\bf B304} (1988) 1.}

\lref\Dine{M. Dine, N. Seiberg, X.G. Wen and E. Witten, {\it Nonperturbative
Effects on the string world-sheet,2}, Nucl. Phys {\bf B289} (1987) 319.}

\lref\Lukas{A. Lukas and K.S. Stelle,
{\it Heterotic anomaly cancellation in five dimensionas},
JHEP {\bf 01} (2000) 010, hep-th/9911156.}

\lref\blume{R. Blumenhagen, G. Honecker and T. Weigand, {\it Lopp-corrected
compactifications of the heterotic string with line bundles}, JHEP {\bf
0506} (2005) 020, hep-th/0504232.}

\lref\AH{B. Andreas and D. Hern\'andez Ruip\'erez,
{\it U(n) Vector Bundles on Calabi-Yau 
Threefolds for String Theory Compactifications}, 
Adv. Theor. Math. Phys. {\bf 9} (2006) 253, hep-th/0410170.}

\lref\FMW{R. Friedman, J. Morgan and E. Witten, 
{\it Vector bundles and $F$-theory},
Comm. Math. Phys. {\bf 187} (1997) 679, hep-th/9701162.}

\lref\FMWII{R. Friedman, J. Morgan and E. Witten,
{\it Vector bundles over elliptic fibrations}, J. Algebraic Geom., 
{\bf 8} (1999), pp.~279--401,  alg-geom/9709029.}

\lref\W{E. Witten, 
{\it New issues in manifolds of $SU(3)$ holonomy}, Nucl. Phys.
{\bf B268} (1986) 79.}

\lref\C{G. Curio, 
{\it Chiral matter and transitions in heterotic string models},
Phys.Lett. {\bf B435} (1998) 39, hep-th/9803224.} 

\lref\ov{R. Donagi, B.A. Ovrut, T. Pantev and D. Waldram,
{\it  Standard Models from Heterotic M-theory},
Adv.Theor.Math.Phys. {\bf 5} (2002) 93,
hep-th/9912208.}

\lref\ovrut{R. Donagi, B.A. Ovrut, T. Pantev and D. Waldram,
{\it Spectral involutions on rational elliptic surfaces},
Adv.Theor.Math.Phys. {\bf 5} (2002) 499,
math.AG/0008011.
{\it Standard-model bundles},
Adv.Theor.Math.Phys. {\bf 5} (2002) 563,
math.AG/0008010.
{\it Standard-Model Bundles on Non-Simply Connected Calabi--Yau Threefolds},
JHEP {\bf 0108} (2001) 053,
hep-th/0008008.}

\lref\ron{V. Bouchard and R. Donagi, {\it  An SU(5) Heterotic Standard Model},
Phys.Lett. {\bf B633} (2006) 783, hep-th/0512149.}

\lref\UhYa{K.~Uhlenbeck and S.-T. Yau, {\it On the existence of
Hermitian Yang-Mills connections in stable vector bundles}, Comm. Pure
Appl. Math. {\bf 39} (1986), pp.~S257--S293,
Frontiers of the mathematical sciences: 1985 (New York, 1985).}

\lref\Donald{Donaldson, Proc. London Math. Soc. 3, 1 (1985).}

\lref\Friedmanbook{R. Friedman, 
{\it Algebraic Surfaces and Holomorphic Vector Bundles},
Springer, Universitext (1998).}

\lref\kobay{S. Kobayashi, {\it Differential Geometry of Complex 
Vector Bundles}, Princeton University Press 1987.}

\lref\Radu{R. Tatar and T. Watari, {\it Proton decay, Yukawa couplings 
and underlying gauge symmetry in string theory}, 
Nucl.Phys. {\bf B747} (2006) 212-265, hep-th/0602238.}

\lref\Braun{V.~Braun, Y.~H.~He, B.~A.~Ovrut and T.~Pantev,
  {\it Vector bundle extensions, sheaf cohomology, and the heterotic standard
  model,}
  Adv.\ Theor.\ Math.\ Phys.\  {\bf 10} (2006) 4, hep-th/0505041.}

\lref\timoone{R.~Blumenhagen, S.~Moster and T.~Weigand,
 {\it Heterotic GUT and standard model vacua from simply connected Calabi-Yau
  manifolds,}
  Nucl.\ Phys.\  B {\bf 751}, 186 (2006), hep-th/0603015.}
 
\lref\timotwo{R.~Blumenhagen, S.~Moster, R.~Reinbacher and T.~Weigand,
{\it Massless spectra of three generation U(N) heterotic string vacua,''}, hep-th/0612039}


\newsec{Introduction}

The goal of the present paper is to get a (supersymmetric) 
phenomenological spectrum from the $E_8\times E_8$
heterotic string on a Calabi-Yau space $X$. More precisely, 
one wants to construct a model leading in four dimensions to the
gauge group and net chiral matter content of the Standard Model.
The individual number of generations
and anti-generations and the number of Higgs multiplets 
will be investigated elsewhere.

A common method to get the Standard Model gauge group $G_{SM}$ 
is to have first a GUT gauge group $H=SU(5)$ 
from embedding a vector bundle of structure group $G=SU(5)$
into the first $E_8$. Then, if $\pi_1(X)={\bf Z_2}$, one uses a Wilson line  
to break $H$ to $G_{SM}$.

We will choose $X$ to be elliptically fibered over $B={\bf F_k}$; 
$X$ is smooth for $k=0,1,2$. We choose a specific fibration type
(the $B$-fibration) which has two sections $\si_1, \si_2$. On this $X$ we find
a free involution $\tau_X$ (as needed because $\pi_1(X)=0$) for $k$ even.
We find invariant bundles of $\pm 6 $ generations
to get the Standard Model on the quotient $X'=X/{\bf Z_2}$.

The vector bundle on $X$ will be constructed using the method of
bundle extensions investigated in \refs{\extpaper}, cf.~also \refs{\Braun}.
More precisely, we 
consider extension bundles of rank $n+m$ defined by 
\eqn\extensinm{0\to \pi^*E_n\ox \cO_X(-mD)\to V_{n+m}
\to \pi^*E_m\ox \cO_X(nD)\to 0}
with $D=x\Si+\pi^*\al$ (where $\Si:=\si_1 + \si_2$)
and $E_i$ vector bundle on the base of vanishing first Chern class 
which are stable with respect to a K\"ahler class $H$ on $B$. 
We will show that $V_{n+m}$ is stable 
with respect to the K\"ahler class $J=z\Si+\pi^*H$ for the real number 
$z>0$ in a suitable range.
One finds that the generation number $N_{gen}=c_3(V_{n+m})/2$
is proportional to $x$
which one therefore has to choose to be non-zero
(the actual physical net number of Standard Model generations
is computed downstairs on $X'$ and is $N_{gen}/2$).

In order that $V_{n+m}$ qualifies 
as a physical gauge bundle it has to satisfy 
the anomaly constraint that requires (cf. below)
$W=w_B\Si +a_f F =c_2(X)-c_2(V_{n+m})$ to be an effective class
(we put here the trivial bundle in the hidden sector). So in particular
\eqn\wbre{w_B=\Big(6-{1\over 2}nm(n+m)x^2\Big)c_1+nm(n+m)x\al}
has to be an effective curve class in $B$. As we are interested in rank five 
vector bundles we find for $(n,m)$ 
with $(3, 2)$ or $(4,1)$ the first term in \wbre\ to be negative. 
Moreover, stability of $V_{n+m}$ 
requires for $x\neq 0$ that $ x\al H\leq 0$. This implies that the last term 
in \wbre\ also contributes negatively 
(meaning here: it can not be a non-zero effective class). 
So the construction \extensinm\
favors physical gauge bundles of rank four (or less) 
of type $(n,m)=(2,2)$ 
(cf.~below; the case $(3,1)$ is also ruled out by the same $w_B$ 
not effective argument). 

An invariant model on the cover space $X$ with GUT gauge group $SU(5)$ arises 
actually not from an $SU(5)$ bundle but from
an $SU(4)\times U(1)$ bundle (of $\pm 6$ net generations)
\eqn\vfive{
V_5=V_4\otimes \cO_X(-\pi^*\beta)\oplus \cO_X(4\pi^*\beta)}
The $U(1)$ of the commutator subgroup $SU(5)\times U(1)$ respresenting
the unbroken gauge group in the first $E_8$ is for $\al \beta \neq 0$
anomalous, becomes massive
and decouples from the low-energy spectrum. 
$V_5$ is poly-stable and invariant; actually $\beta = \pm (1, -1)$.

The invariant rank 4 bundle $V_4$ of $c_1(V_4)=0$ 
is defined by a non-split extension
\eqn\ebun{0\to \pi^*E_1\ox \cO_X(-D)\to V_4\to \pi^*E_2\ox \cO_X(D)\to 0}
Here and from now on $E_1$ and $E_2$ refer to bundles of rank two on the base.

We will show that the extension can be chosen non-split and $V_4$ 
to be invariant.
A crucial part of the argument will be to show the stability of $V_4$ 
with respect to a K\"ahler form $J=z\Si +\pi^* H$. 
For a technical reason the K\"ahler class $H$ on the base
has to be chosen to be proportional to $c_1:=c_1(B)$; so we 
will work finally over ${\bf F_0 = P^1\times P^1}$.

Assuming that the bundles $E_i$ on the base are chosen to be invariant
under the base part $\tau_B$ of the involution $\tau_X$
the pull-back bundles turn out to be invariant as well.
The line bundle twist $\cO_X(D)$ can be chosen invariant 
as can the extension bundle $V_5$.

So the original problem to construct a suitable bundle on $X$ 
is reduced to construct invariant bundles $E_i$ on the base
of suitable instanton numbers $k_i=c_2(E_i)$
(to get $N_{gen}=\pm 6$). Examples of appropriate base bundles 
are given in the appendix. Thus we have achieved our goal to
construct a heterotic Standard Model compactification.
For other heterotic derivations of the Standard Model 
via an intermediate $SU(5)$ GUT group cf.~\refs{\ovrut, \ron}.
\noindent
The physical generation number $N_{gen}^{phys}=N_{gen}/2$
(downstairs on $X'=X/{\bf Z_2}$) 
is given by 
\eqn\mainform{N_{gen}^{phys}=k_1-k_2}
The $E_i$ are two stable, invariant  rank two vector bundles on $B$ 
of $c_1(E_i)=0$ and $c_2(E_i)=k_i$. 
We find that $D$ is given by the invariant divisor $D=\Si +\pi^*\al$ 
where $\al$ is $(-2,-2)$ or $(-1,-1)$.
The list of applicable instanton numbers
$(k_1, k_2)$ for the various  choices of $\al$ 
is given in table 1 (for $\beta=(1, -1)$; for the negative of that
one just has to interchange again the $k_i$).
These data-sets fulfill all necessary conditions for the existence
of a non-split extension, stability, DUY-equation,
fivebrane effectivity and generation 
number.

There is a fundamental alternative in this construction:
one can either, as described above, cancel unwanted exotic matter 
(apart from a wellcomed right-handed neutrino $\nu_R$)
produced by the 
decomposition of ${\bf 248}$ under $SU(4)\x SU(5) \x U(1)_X$; the condition
is $\al \beta =0$, thus one keeps the $U(1)$ which is now not anomalous.
Or one keeps that matter but avoids the $U(1)$; 
then the solutions are given in appendix $D$.
(or their reflections (under $(p,q)\to (q,p)$),
when at the same time the numbers $k_1$ and $k_2$ are interchanged).

\bigskip
\noindent
{\it Relation to Previous Bundle Constructions}

To put the construction presented in this paper into perspective, 
let us indicate how it
has arisen as culmination of previous investigations
\refs{\BfibreI\BfibreII\BfibreIII\extpaper-\physpaper} 
along similar lines which incorporated different subsets of the
whole procedure. 

Attempts to get a (supersymmetric) phenomenological spectrum with gauge
group $G_{SM}$ and chiral matter content of the Standard model 
from the $E_8\times E_8$ heterotic string on a Calabi-Yau space $X$
started with embedding the spin connection in the gauge connection giving an
unbroken $E_6$ (times a hidden $E_8$ coupling only gravitationally). 
More generally \refs{\W}, one can instead of the tangent bundle embed an 
$G=SU(n)$ bundle for $n=4$ or $5$,
leading to unbroken $H=SO(10)$ or $SU(5)$ of even greater phenomenological
interest. A concrete description of vector bundles
on a general Calabi-Yau space $X$ (not given via projective embedding)
was made in \refs{\FMW} for the case that $X$ has an elliptic fibration
$\pi:X \ra B$. The net generation
number for these bundles was computed in \refs{\A,\C}.
It was soon realized that the only elliptically fibered $X$
that has non-trivial fundamental group
has the Enriques surface as base and leads to generation number zero.

Therefore the following indirect strategy had to be employed.
If there is an freely acting group $\G$ on the usually simply-connected $X$,
one can work on $X'=X/\G$ with $\pi_1(X')=\G$ allowing a further breaking
of $H$ by turning on Wilson lines. 
This was achieved in \refs{\BfibreI} in a general way by using a non-standard
elliptic fiber (the $B$-fibration) which leads to an elliptic fibration of $X$
having two sections $\si_1$ and $\si_2$. 
This led to a free involution $\tau_X$ with the required properties. 

The necessary invariance of the bundle
(so that it descends to the quotient) was checked first \refs{\BfibreI} 
only on the level
of cohomological invariants (cf.~\refs{\ov} for a similar procedure).
Then the related action on the spectral parameters defining the bundle 
and the corresponding invariance was investigated \refs{\BfibreII}.
Due to an ensuing integrality problem by a factor $1/2$ (essentially the
question whether $\Si = \si_1+\si_2$ can be assumed to restrict to
an even class on the spectral surface)
the spectral bundle construction itself was adapted more
properly to the case of the $B$-fibration \refs{\BfibreIII}.

Although the invariance problem was solved, as a side effect
the necessity of a structure group of even rank emerged and also
an even number of Standard Model net generations. The first issue is
overcome by using an $SU(4)$ bundle $V_4$ and embedding the $U(4)$ bundle
(emerging after a line bundle twist)
\refs{\AH}
in $E_8$ via $V_4\otimes \cL \oplus \cL^{-4}$.
The rank $5$ bundle arising then has to be poly-stable (which in case that
$c_1(\cL)=y\Si + \pi^*\beta$ has non-zero $y$ leads to the necessity of 
including one-loop effects \refs{\BfibreIII}). Because of the second problem 
the class of bundles was enlarged \refs{\extpaper, \physpaper} 
by considering also non-split extensions
of line bundles by $SU(4)$ bundles, 
the latter chosen to be pulled back from the base rather than being spectral. 
There the case of the $A$-fibration
was considered, leading to GUT models, and the Enriques base giving
the Standard Model gauge group; the latter case leads 
to problems when including the condition of effectivity of the fivebrane class
from anomaly cancellation.

Using extensions in the case of the $B$-fibration for spectral
or pull-back bundles is possible, but leads again to problems with 
the effectivity of the fivebrane class.
Only when $y=0$ did these problems disappear, but this also 
suppresses the chiral generations. Therefore this generation number
has to emerge from a different parameter which in fact should be present
already in $V_4$. Hence we employ the more general
bundle construction of an extension of a bundle of higher rank 
(no longer a line bundle) by another bundle, already touched upon in
\refs{\extpaper, \physpaper}. This is done here for the case 
of two bundles of rank two, which themselves arise as pullbacks twisted
by a line bundle (having $x\neq 0$) as described above. 
Then here an extension of a line bundle 
by such a rank 4 bundle is considered.

\bigskip
In {\it section 2} we set the stage and to establish some notation.
We recall some generalities of the bundle
construction concerning extensions, stability, the physical constraints
of effectivity of the fivebrane class and the phenomenological net number
of chiral matter generations.
In {\it section 3} we review the Calabi-Yau spaces with $B$-fibration,
which have two sections and admit a free involution $\tau_X$.
In {\it section 4} we construct the extension described above, show 
that it can be chosen non-split and that $V_5$ is stable. 
In {\it section 5} we describe the way how to get the rank $5$ bundle
from the rank $4$ bundle and various question connected with this procedure.
In {\it section 6} we collect all constraints and present some solutions. 
In {\it section 7} we collect our conclusions.
In the {\it appendix} we give some useful cohomological formulae and examples 
of stable and invariant bundles $E_i$ on the base and specify the 
numerical constraints on the instanton numbers. Further 
we describe a general argument how to count the number of invariant bundles.

\newsec{Bundles and Physical Constraints}

We begin with some general remarks on the constraints on the bundles 
used in a heterotic compactification.
These concern first the equations of motion of the underlying string theory.
Thanks to the work of Donaldson, Uhlenbeck and Yau this can be translated
to the mathematical condition of stability. Then we move on
to the special case that the bundle $V$ arises as an extension 
of other bundles. In this case one gets immediately two 
necessary conditions from the stability of $V$, in particular in our case
a non-split condition for the extension arises.

Up to this point the physical conditions are the same 
as a pure mathematical investigation would pose, namely the
requirement of stability.
After this the physical investigation proceeds
to pose further requirements: first, one condition of 
physical consistency (anomaly cancellation, this comes down to 
the effectivity of the five-brane class); then a phenomenological
requirement on the number of Standard model net generations is posed.

\subsec{Stability}

The main order in constructing a heterotic compactification is to solve 
the physical equations of motion. For the underlying space this can be reduced 
to the topological question of constructing a Calabi-Yau space.
For the bundle sector one reduces the K\"ahler-Yang-Mills equations
for a $G$-valoued connection, via the Donaldson-Uhlenbeck-Yau (DUY) theorem, 
to the construction of a holomorphic vector bundle which has to be stable.
Like the Calabi-Yau condition on the underlying space $X$, the
holomorphicity and stability of the vector bundle $V$ are direct
consequences of the required four-dimensional supersymmetry. The demand is
that a connection $A$ on $V$ has to satisfy the DUY equation
\eqn\hym{F_A^{2,0}=F_A^{0,2}=0, \quad F_A^{1,1}\wedge J^{2}=0}
The first equation implies the holomorphicity of $V$; the second equation is
the  Hermitian-Yang-Mills (HYM) equation $F_A^{1,1}\wedge J^{n-1}=c\cdot
I_F\cdot J^n$ for $n=3$ with $c\in {\bf C}$ vanishing. The latter
has, after taking the trace and integrating, the integrability condition
\refs{\UhYa, \Donald}
\eqn\duy{\int_X c_1(V)\wedge J^2=0}
This necessary condition becomes sufficient for the existence of a unique 
solution if $V$ is stable (or, more generally, polystable,
i.e., a sum of stable bundles with the same slope).
Stability of $V$ (with respect to $J$) means $\mu_J(V')< \mu(V)$
for all coherent subsheafes $V'$ of $V$ of $rk \, V' \neq 0, rk \, V$
(it suffices to test the $V'$ with $V/V'$ torsion-free,
cf.~Ch.~4, Lemma 5 \refs{\Friedmanbook}).
Here $\mu(V)= {1 \over rk\, V} \int c_1(V) J^2$ is the slope of $V$ 
with respect to $J$.

\subsec{Extension Bundles and the Non-Split Condition}

For a zero-slope bundle $V$ constructed as an extension 
(with $U$ and $W$ stable)
\eqn\nonsplit{0 \to U \to V \to W \to 0}
one finds two immediate conditions which are necessary for stability

i) $\mu(U)<0$ 

ii) the $W$ of $\mu(W)>0$ is not a 
subbundle of $V$, i.e., the extension \nonsplit\ is non-split

\noindent
The first condition comes down in our case 
\eqn\positslope{DJ^2=2x(h-z)^2c_1^2+2z(2h-z)\al c_1\; > \; 0}
The non-split condition can be expressed
as $Ext^1(W,U)=H^1(X, W^*\ox U)\neq 0$.

\subsec{Anomaly Constraint and Net Generation Number}

Anomaly cancellation forces the three-form field strength $H$ to satisfy 
$dH=trR\wedge R-TrF\wedge F$
where $R$ and $F$ are the curvature forms of the spin connection 
on $X$ and the gauge connection on $V$. This gives the topological condition
$c_2(TX)=c_2(V)$.
The inclusion of (magnetic) five-branes changes the topological constraint 
\FMW\ on the gauge bundle $V$ by contributing a source term 
to the Bianchi identity for the three-form $H$
$dH=trR\wedge R-TrF\wedge F-n_5\sum_{\rm five-branes}\delta_5^{(4)}$.
The current $\delta_5^{(4)}$ integrates to one in the 
direction transverse to a five-brane of class $[W]$. 
Integration over a four-cycle in $X$ gives $c_2(TX)=c_2(V)+[W]$.
Supersymmetry requires that five-branes are wrapped on holomorphic curves
and $[W]$ has to be an effective class. 
So the effectivity of $[W]$ constraints the choice of vector bundles $V$.  
In the elliptically fibered case $H^4(X)$ decomposes as
\eqn\ellipdecomp{
H^4(X)=\si_1\, H^2(B) \oplus  \si_2\, H^2(B) \oplus \pi^* H^4(B)}
Actually, $c_2(X)$ (cf.~below) and $c_2(V)$ (by the invariance requirement)
lie in the symmetric subspace $\Si \, H^2(B) \oplus \pi^* H^4(B)$
where $\Si=\si_1 + \si_2$. The effectivity condition for the class
\eqn\fivebr{W=w_B \Si + a_f F}
becomes (here $w_B\geq 0$ means that the class is effective)
\eqn\anogenn{w_B\geq 0\; , \;\;\;a_f\geq 0}

In general, the decomposition of the ten-dimensional Dirac operator 
with values in $V$ shows that massless four-dimensional fermions are in one 
to one correspondence with zero modes of the Dirac operator $D_V$ on $X$
whose index is
${\rm index}(D_V)=\sum_{i=0}^3(-1)^k {\rm dim}H^k(X,V)=\int_X {Td}(X)ch(V)$.
For stable vector bundles one has $H^0(X,V)=H^3(X,V)=0$ and so
${\rm dim}\, H^2(X,V)-{\rm dim}\, H^1(X,V)={1\over 2}\int_X c_3(V)$.
For the net number of chiral matter generations 
one gets with  
\eqn\nge{N_{gen}=h^1(X, V_5^*)-h^1(X, V_5)=\int_X ch(V_5)Td(X)
=\int_X {c_3(V_5)\over 2}}
which we want to equal $\pm 6$ in order to get downstairs on $X'=X/{\bf Z_2}$ 
the $\pm 3$ phenomenological net generations of Standard Model fermions.

\newsec{Review of the Elliptic Calabi-Yau Space with Two Sections}

We consider a Calabi-Yau threefold $X$, 
elliptically fibered over a Hirzebruch surface ${\bf F}_m$, whose 
generic fiber is described by the so-called $B$-fiber ${\bf P}_ {1,2,1}(4)$
instead of the usual $A$-fiber ${\bf P}_{2,3,1}(6)$ (the subscripts 
indicate the weights of $x,y,z$). $X$ is given by a generalized 
Weierstrass equation which embeds $X$ in a weighted projective space 
bundle over ${\bf F_m}$
\eqn\weier{y^2+x^4+a_2x^2z^2+b_3xz^3+c_4z^4=0}
where $x,y,z$ and $a,b,c$ are sections of $K_B^{-i}$ with $i=1,2,0$ 
and $i=2,3,4$, respectively.

$X$ admits two cohomologically inequivalent section $\si_1,\si_2$. 
For this consider \weier\ at the locus $z=0$, i.e.,
$y^2=x^4 $ (after $y\to iy$).
One finds 8 solutions which constitute the two equivalence classes 
$ (x,y,z)=(1,\pm 1,0)$ in ${\bf P}_{1,2,1}$. We choose $y=+1$, 
corresponding to the section $\si_1$, as zero in the group law, 
while the other one can be brought, for special points in the moduli space, 
to a half-division point (in the group law) leading to the shift-involution. 
Let us keep on record the relation of divisors 
(with $\si_i:=\si_i(B)$, $i=1,2$) 
\eqn\sma{(z)=\Si:=\si_1+\si_2, \ \ \  \si_1\cdot \si_2=0}

The fibration structures leads in the following way to the 
cohomological data of $Z$ (unspecified Chern classes like $c_1$ 
refer to the base $B$; further we write $c_i=\pi^*c_i(B)$). 
 
As noted $z,x,y$ can be thought of as homogeneous coordinates on a
${\bf P}_{1,2,1}$ bundle $W$, i.e. as sections of 
line bundles ${\cal O}(1), {\cal O}(1)\otimes {\cal L}$ and 
${\cal O}(1)^2\otimes {\cal L}^2$ whose first Chern classes 
are given by $r$, $r+c_1$, $2r+2c_1$ with $c_1({\cal O}(1))=r$.
The cohomology ring of $W$ is generated by $r$ with the relation 
$r(r+c_1)(2r+2c_1)=0$ expressing the fact that $z,x,y$ have no common zeros. 
As the $B$-model is defined by the vanishing 
of a section of ${\cal O}(1)^4\otimes {\cal L}^4$, which is a line bundle over 
$W$ with first Chern class $4(r+c_1)$,
the restriction from $W$ to $X$ is effected by multiplying by this 
Chern class, so that $c(W)=(1+4r+4c_1)c(X)$. One can then simplify
$r(r+c_1)(2r+2c_1)=0$ to $r(r+c_1)=0$ in the cohomology ring of $X$ 
and finds 
\eqn\total{c(X)=c(B){(1+r)(1+r+c_1)(1+2r+2c_1)\over {1+4r+4c_1}}}
With $r^2=-rc_1$ and the class 
$r=\sigma_1+\sigma_2$ (as $z=0$ implies $y^2=x^4$ giving
$(x,y)=(i,1)$ and $(i,-1)$) 
of the divisor $(z=0)$ of the section $z$ of the line bundle ${\cal O}(1)$ 
we find
\eqn\charclass{c_2(X)\; = \; \pi^*c_2+6\Si \pi^*c_1 +5\pi^*c_1^2\; , \;\;\;
c_3(X)\; =\; -36 \pi^*c_1^2}

From the weights $a_2$, $b_3$ and $c_4$ of the defining equation 
one gets $5^2+7^2+9^2-3-3-1=148$ complexe structure deformations over 
${\bf F_0}$. This is consistent with the Euler number and the 
$h^{1,1} (X)=4$ K\"ahler classes
\eqn\hodge{h^{1,1}(X)=4, \ \ \  h^{2,1}(X)=148, \ \ {\rm and}\ \  e(X)=-288}
For later use let us also note the adjunction relations 
\eqn\adjrel{\si_i^2=-\pi^*c_1 \si_i\, , \;\;\;\;\; \Si^2=-\pi^*c_1\Si}

\bigskip\noindent{\it The K\"ahler Cone}

For the base $B$ being 
given by a Hirzebruch surface ${\bf F_m}$ (with $m=0,2$) 
$H^2(B, {\bf Z})$ generated by the effective base and fiber classes $b$ 
and $f$ (with intersection relations $b^2=-m$, $b\cdot f=1$ and $f^2=0$).
Obviously, these two classes represent actual curves. The effective cone
(non-negative linear combinations of classes of actual curves)
is given by the condition $p\geq 0, q\geq 0$ on $\rho =pb+qf$;
this we denote by $\rho \geq 0$. The K\"ahler cone ${\cal C}_B$ of $B$
(where $\rho \in {\cal C}_B$ means $\rho \zeta > 0$ for all actual 
curves of classes $\zeta$ or equivalently $\rho b>0, \rho f>0$) 
is given by ${\cal C}_B=\{t_1 b^+ + t_2 f| t_i >0 \}$ (with $b^+=b+mf$).
For example on ${\bf F_2}$ one has $c_1\notin {\cal C}_B$ as $c_1b=0$.

Let $J=x_1\si_1+x_2\si_2+\pi^* H$ be an element
in the K\"ahler cone ${\cal C}_X$ ($H \in {\cal C}_B$).
Demanding that its intersections with the curves $F$ and $\si_i \alpha$
are non-negative amounts to $x_1+x_2 >  0$ and $(H-x_ic_1) \alpha >  0$.
Similarly intersecting $J^2$ with $\si_i$ and $\alpha$ gives the conditions
$(H-x_ic_1)^2 >  0$ and
$( 2 \sum x_i \; H - \sum x_i^2 \; c_1 ) \alpha >  0$.
Integrating $J^3$ gives
$\sum x_i \, (H-x_i c_1)^2 + ( 2 \sum x_i \; H - \sum x_i^2 \; c_1) H >  0$.
From this one gets the condition for $J$ to be ample (positive)
\eqn\KK{J=x_1\si_1+x_2\si_2+\pi^*H\in {\cal C}_X \;\;\; \Longleftrightarrow 
\;\;\; x_1+x_2>0\;\;\; , \;\; H-x_ic_1 \in  {\cal C}_B }
Concretely we will choose $J=z\Si + \pi^* hc_1$ giving the condition
$0 < z < h$. Below we will restrict to the case $B={\bf F_0}$
as we will have to use the fact that $c_1\in {\cal C}_B$.

\subsec{Existence of a Free ${\bf Z}_2$ Operation}

We give a free involution $\tau_X$ on $X$ which leaves the holomorphic 
three-form invariant; then $X'=X/{\tau_X}$ is a smooth Calabi-Yau.
We assume $\tau_X$ compatible with the fibration, i.e., we assume the
existence of an involution $\tau_B$ on the base $B$ with $\tau_B\cdot
\pi=\pi\cdot \tau_X$.

We will choose for $\tau_B$ the following operation in local (affine)
coordinates
\eqn\Binvo{b=(z_1, z_2)\;\;\; \buildrel{\tau_B}\over \longrightarrow \;\;\;
-b=\tau_B(b)=(-z_1, -z_2)}

To define $\tau_X$ one combines $\tau_B$
with an operation on the fibers (cf.~\refs{\BfibreI, \BfibreII}).
A free involution on a smooth elliptic curve 
is given by translation by a half-division point. Such an object has to exist 
globally; this is the reason we have chosen the $B$-fibration where $X$ 
possesses a second section. If we would tune $\si_2(b)\in E_b$
to be a half-division point the condition $b_3=0$ would ensue and $X$
would become singular. Therefore this idea has to be enhanced.
Furthermore, even for a $B$-fibered $X$ those fibers
lying over the discriminant locus in the base will be singular where
the freeness of the shift might be lost. As the fixed point locus of $\tau_B$
is a finite set of (four) points we can assume that it is disjoint
from the discriminant locus 
(so points in the singular fibers are still not fixed points of $\tau_X$).

One finds \refs{\BfibreI} as $\tau_X$ over ${\bf F_m}$ 
with $m$ even (i.e., $m$ being $0$ or $2$) the free involution
\eqn\iota{(z_1, z_2 \, ; x,y,z)\buildrel{\tau_X}\over
\longrightarrow (-z_1, -z_2 \, ; -x,-y,z)}
This exchanges the points $\si_1(b)=(b \, ; 1,1,0)$ and 
$\si_2(-b)=(-b\, ;  1,-1,0)$
between the fibers $E_b$ and $E_{-b}=E_{\tau_B(b)}$; in ${\bf  P}_{1,2,1}$
the sign in the $x$-coordinate can be scaled away here in contrast to the
sign in the $y$-coordinate.  As indicated above an involution like in 
\iota\ could not exist on the fiber alone, i.e.
as a map $(x,y,z)\lra (-x, -y, z)$, because this would 
force one, from \weier, to the locus $b_3=0$ where $X$ becomes singular
(so only then is this defined on the fiber and so, being a free
involution, a shift by a half-division point).
But it can exist combined with the base involution $\tau_B$ on a
subspace of the moduli space where the generic member is still
smooth: from \weier\ the coefficient functions should transform
under $\tau_X$ as $a_2^+, b_3^-, c_4^+$, so over $ {\bf F_0}$
only monomials $z_1^pz_2^q$ within $b_{6,6}$ with $p +q$ even 
(in $a_{4,4}$ and $c_{8,8}$ with $p+q$ odd) are forbidden.  
So the number of deformations drops to $h^{2,1}(X) = (5^2
+1)/2+(7^2-1)/2+(9^2+1)/2-1-1-1=75$. The discriminant remains generic
as enough terms in $a,b,c$ survive, so $Z$ is still smooth, 
cf.~\refs{\BfibreI}. The Hodge numbers $(4,148)$ and $(3, 75)$ of $X$ and $X'$
show that indeed $e(X')=e(X)/2$
($X'$ has lost one divisor as the two sections are identified).

\newsec{Stable $SU(4)$ Bundles on $X$}
 
Let $E_1$ and $E_2$ be two stable rank two vector bundles on $B$
of $c_1(E_i)=0$ and $c_2(E_i)=k_i$.
We consider the extension defining our rank four bundle 
(with $D=x\Si+\pi^*\al$)
\eqn\rankf{0\to \pi^*E_1\ox \cO_X(-D)\to V_4\to \pi^*E_2\ox \cO_X(D)\to 0}

\subsec{Stability of $\pi^*E_i$}

We prove that $\pi^*E$ is stable on $X$ with respect to a
$J=z\Si +\pi^*H$ in the K\"ahler cone 
${\cC}_X$ $($i.e. $H-zc_1\in {\cC}_B$ \refs{\BfibreIII}, so $z<h)$ 
if $E$ of $c_1(E)=0$ is stable on $B$ with respect to $H=hc_1$.
Thus, for $c_1\in {\cal C}_B$, we assume from now on that $B={\bf F_0}$. 
Following Lemma 5.1 \extpaper\
let $\F$ be a subsheaf of $\pi^*E$ where we can assume that $\pi^*E/\F$ 
is torsion free (cf.~Ch.~4, Lemma 5 \refs{\Friedmanbook});
we have $0\to \F\vert_{\si_i}\to E$ and $c_1(\F\vert_{\si_i})H<0$. 
Similarly we get $0\to \F\vert_{F}\to \cO^r_F$
thus $deg(\F\vert_{F})\leq 0$ as $\cO^r_F$ is semistable (where $r:=rk(E)$). 
Then for $H-zc_1\in {\cC}_B$ and
$c_1(\F) = -A_1\si_1-A_1'\si_2 + \pi^*\la$ with $(A_1+A_1')\geq 0$ and 
$\la H \leq (A_1c_1+\la)H< 0$ (the same holds for $A_1'$)
\eqn\slopeF{c_1(\F)J^2 = -(A_1+A_1')(H-zc_1)^2 + 2z ( 2H-zc_1)\la < 0.}

\subsec{Non-Split Condition}

We derive a condition such that the extension can be chosen 
non-split (as necessary for  $V_4$ stable, cf.~\nonsplit). 
An extension can be chosen non-split if (with $\cE:=E_1\ox E_2^*$)
\eqn\nspco{Ext^1\Big(\pi^*E_2\ox \cO_X(D), \pi^*E_1\ox \cO_X(-D)\Big)
=H^1\Big(X, \pi^*\cE \ox \cO_X(-2D)\Big)\neq 0}

In the following we will discuss the case with $x> 0$. One has 
(let $y=2x$)
\eqn\negxsecond{x>0\, : \;\;\;
\pi_*\cO_X(-y\Si)=0 \, , \;\;\;\; R^1\pi_*\cO_X(-2x\Si)=
\cO_B\oplus K_B\oplus 2K_B^{-1}\oplus\dots \oplus 2K_B^{1-y}}
(cf.~appendix).
The Leray spectral sequence yields then the isomorphism 
\eqn\lerayn{H^1\Big(X, \pi^*\cE \ox\cO_X(-2D)\Big)
\simeq H^0\Big(B, \cE \ox \cO_B(-2\al)\ox R^1\pi_*\cO_X(-y\Si)\Big)}
(cf.~appendix).
Because of \negxsecond\ it suffices to show that one of the terms
in the corresponding decomposition of \lerayn\ is non-vanishing.
We give sufficient conditions 
for $H^0(B, \cE\ox \cO_B(-2\al)\otimes K_B^{1-y})\neq 0$.
For this we will compute the expression
\eqn\hrrind{\chi\Big(B,  \cE\otimes \cO_B(-2\al)\otimes K_B^{1-y}\Big)
=\sum_{i=0}^2(-1)^i{\rm dim}
H^i\Big(B, \cE\otimes \cO_B(-2\al)\otimes K_B^{1-y}\Big)}

So the sequence defining $V_4$ can be chosen non-split
if $\chi(B,  \cE\otimes \cO_B(-2\al)\otimes K_B^{1-y})>0$ 
and $H^2(B, \cE\otimes \cO_B(-2\al)\otimes K_B^{1-y})
=H^0\Big(B, \cE^*\otimes \cO_B(2\al)\otimes K_B^{y}\Big)^*=0$.
Now $H^0(B, \cE^*\otimes \cO_B(2\al)\otimes K_B^{y})=0$ 
if $\mu(\cE^*\otimes \cO_B(2\al)\otimes K_B^{y})=12(2\al-yc_1)z(2h-z)c_1<0$ 
(a section gives a slope zero subbundle and 
$\cE^*$ is semistable (\refs{\kobay}, Thm. 10.16) 
of zero-slope), i.e., if $2\al-yc_1=2(\al-xc_1)<0$. 
The Hirzebruch-Riemann-Roch theorem gives the condition 
\eqn\index{\eqalign{0\; &< \; 
{1\over 2}\chi\Big(B, \cE\otimes \cO_B(-2\al)\otimes K_B^{1-y}\Big)
={1\over 2}\int_B ch(\cE)ch(\cO_B(-2\al))ch(K_B^{1-y})Td(B)\cr
&=\; 2+\Big((y-1)+(y-1)^2\Big)c_1^2+4\al^2
-2\Big(1+2(y-1)\Big)\al c_1-\Big(k_1+k_2\Big)}}
So $\al - x c_1<0$ and \index\ are sufficient conditions 
for the existence of a non-split extension.

The existence of an {\it invariant} extension bundle $V_4$ 
follows by the same arguments as on $B$: 
the pull-back bundles $\pi^*E_i$ are $\tau_X$-invariant
as the $E_i$ are chosen $\tau_B$-invariant
(the action in the elliptic fiber coordinates is ineffective here);
the divisor $\Si$ is invariant and $\pi^* \al$ can be chosen
invariant as $\al$ itself can from the selection of monomials argument.

\subsec{Stability of the Rank 4 Extension}

Having shown that \rankf\ can be chosen non-split for $x>0$ 
we give the stability proof along the lines of \extpaper, 
giving a range in the K\"ahler cone such that $V_4$ is stable. Consider 
\eqn\stabEF{\matrix{
  &  &0 & & 0   &  &0 & \cr
  &  &\uparrow &                   &\ua&             & \ua &      &\cr
0 &\to &  P=\bar{P}\ox \cO_X(-D) & \to & V/V'_{r+s}&\to & T
=\bar{T}\ox\cO_X(D)&
\to &0\cr
  &      & \ua &                   &\ua&                    & \ua &      &\cr
0 & \to & \pi^*E_1\ox \cO_X(-D)& \buildrel{i}\over{\to} & V_4 & \buildrel{j}
\over{\to}
 &\pi^*E_2 \ox\cO_X(D)& \to & 0\cr
  &      & \ua &                   &\ua&                    & \ua &      &\cr
0 & \to & F_r\ox \cO_X(-D)& {\to}& V'_{r+s} & {\to} & G_s\ox\cO_X(D) & \to & 
0
\cr
 &      & \ua &                   &\ua&             & \ua &      &\cr
 &  &0 & & 0   &  &0 &}}
where $F_r\ox \cO_X(-D)=i^{-1} V'_{r+s}$ and $G_s\ox \cO_X(D)=j(V'_{r+s})$ 
are of rank $0\leq r\leq 2$ and $0\leq s\leq 2$.
First note thar for a general subsheaf $V'_{r+s}$ of $V$ one has
\eqn\slope{(r+s)\mu(V'_{r+s})=r\mu(F_r)+s\mu(G_s)+(s-r)DJ^2}
To prove stability of $V_4$ we have to show that $\mu(V'_{r+s})<0$ for 
$0\leq r\leq 2$ and $0\leq s\leq 2$ with $0<r+s<4$, so the 
cases $(0,0)$ and $(2,2)$ do not have to be considered. 
The cases $(2,s)$ with $0\leq s<2$ do not have to be considered
as we can assume (cf. Lemma 4.5, \Friedmanbook) that the 
quotient $V_4/V'_{r+s}$ is torsion free, but the quotient
$\pi^* E_1\otimes \cO_X(-D)) / (F_2\otimes \cO_X(-D))$ is a torsion sheaf,
thus zero, i.e. $F_2=\pi^* E_1$ and we assume anyway the necessary condition
$DJ^2>0$ such that $\mu(\pi^* E_1\otimes \cO_X(-D)))<0$ (cf.~the discussion
after \nonsplit).
Note also that  $r=0$ implies $F_r=0$ as $\pi^*E_1$ 
does not have a non-zero torsion subsheaf (same for $s=0$). 

As the pullback bundles are stable we have for $0<r<2$ and $0<s<2$ that
\eqn\slopes{\mu(F_r)<0, \ \ \  \mu(G_s)<0.}
We also have $\mu(F_2)\leq 0$ and $\mu(G_2)\leq 0$ (cf. \Friedmanbook). 
So the cases $(r,s)$ with $(1,1)$ and $(1,0)$ 
are done as already $\mu(V_{r+s}')<0$
and we are left with the cases $(1,2)$, $(0,1)$ and $(0,2)$.
As in \extpaper\ we treat these cases by solving the corresponding 
slope inequalities for $z$ thus determining a range 
in the K\"ahler cone ${\cal C}_X$ where $V_5$ is stable.
Note first that
\eqn\slopc{\eqalign{
r\mu(F_r)&= -(A_1+A_1')(h-z)^2c_1^2+2z(2h-z)\la_1 c_1\cr
s\mu(G_s)&=-(A_2+A'_2)(h-z)^2c_1^2+2z(2h-z)\la_2 c_1\cr
DJ^2&=2x(h-z)^2c_1^2+2z(2h-z)\al c_1}}
where $(A_1+A_1')\geq 0$ and $(A_2+A_2')\geq 0$. 
Further we will assume that $x>0$ and $DJ^2>0$. 
\medskip\noindent{\it The cases $(1,2)$ and $(0,1)$}: As $\mu(G_2)\leq 0$, 
it is sufficient to solve $\mu(F_1)+DJ^2<0$ for $z$. 
As $(A_1+A_1')\geq 0$ we will assume $(A_1+A_1')=0$ as this 
gives the strongest condition. 
Similarly from the stability of $\pi^*E_1$ we have $\la_1c_1<0$ and
the condition will become strongest for $\la_1c_1=-2$.
Let $\zeta:= h-z$ such that $0<\zeta<h$. Then the condition becomes
\eqn\strongcase{x\zeta^2 + (\al c_1 -2)(h^2-\zeta^2)\; < \; 0}
We immediately find the necessary condition $\al c_1 -2<0$ or
\eqn\necesscond{\al c_1 \; \leq \; 0}
Recall that from the non-split condition we assume
$\al-xc_1<0$. 
Then solving the estimated inequality for $z$ we find the bound
(which becomes strongest for $\la c_1 =-2$)
\eqn\boundtt{h^2-\zeta^2>{xc_1^2\over -(\al-xc_1)c_1-\la_1c_1}\; h^2}
A similar discussion leads to \boundtt\ in the $(0,1)$ case 
(then with $\la_2$).  
\medskip\noindent{\it The case $(0,2)$}: here one has to solve 
$2\mu(G_2)+2DJ^2<0$ for $z$. 
As $DJ^2>0$ and $\mu(G_2)\leq 0$ we would have $\mu(V'_{0+2})>0$ if 
$\mu(G_2)=0$ which would destabilize $V_4$ if such a case could actually 
occur. So we have to make sure that 
subsheaves $V'$ of type $(0,2)$ with $\mu(G_2)=0$ do not occur. 
This argument, involving the so-called $f$-map, is given below.

So let us suppose that we can assume $\mu(G_2)<0$; 
so we have to solve for $z$ 
\eqn\zt{2\mu(G_2)+2DJ^2<0}
where we have to treat the cases $A_2+A_2'=0$ with $\la_2 c_1=-2$ 
and $A_2+A_2'=1$ with $\la_2 c_1=0$ (the latter is now possible, cf.~below; we
assume here the minimal values of $A_2+A_2'$ and $\la_2 c_1$ 
corresponding to the largest values of $\mu(G_2)$). 
For $A_2+A_2'=0$ and $\la c_1=-2$ we get the bound
(which is stronger than \boundtt)
\eqn\ztb{h^2-\zeta^2>{2xc_1^2\over -2(\al-xc_1)c_1+2}\; h^2}
and for $A_2+A_2'=1$ and $\la_2 c_1=0$ we get (as $4(\al-xc_1)c_1+c_1^2<0$)
\eqn\ztbb{h^2-\zeta^2>{(4x-1)c_1^2\over -4(\al-xc_1)c_1-c_1^2}\; h^2}
Comparing the above bounds we find the strongest bound is \ztb\
{provided that} $\al c_1<1-4x$ (otherwise it is \ztbb).

In order that \ztb\ can be solved the ratio must be less than $1$, which
again expresses the condition \necesscond.
The condition $DJ^2>0$ imposes the upper bound 
\eqn\uppb{h^2-\zeta^2< {xc_1^2\over -(\al-xc_1)c_1} \; h^2}
thus we find in total the condition ({provided that} $\al c_1<1-4x$, 
cf.~above)
\eqn\window{{xc_1^2\over -(\al-xc_1)c_1+1}\; h^2<h^2-\zeta^2
< {xc_1^2\over -(\al-xc_1)c_1} \; h^2}
So, under the mentioned assumptions, 
$V_4$ is stable with respect to $J=z\Si + h \pi^* c_1$
if \window\ is satisfied for $\zeta=h-z$. 

\bigskip\noindent{\it The $f$-Map Argument}

We still have to make sure that 
subsheaves of type $(0,2)$ with $\mu(G_2)=0$ do not occur. 
More generally, one can pose a condition such that a subsheaf of type $(0,2)$ 
does not exist. 

Let us recall the general argument from \extpaper.
Let $U:=\pi^*E_1\ox \cO_X(-D)$, $W:=\pi^*E_2\ox \cO_X(D)$ 
and $G:=G_2\ox \cO_X(D)$.
A sufficient condition for $V_4$ not to be destabilized by a 
subsheaf $G$ of $W$ is given by
injectivity of the map (which we will call the {\it f-map})
\eqn\fmapone{
{\rm Ext}^1(W, U)\buildrel{f}\over{\to} {\rm Ext}^1(G,U)}
To see this we ask when is it possible that a map $G\to W$ lifts to a map
$G\to V$.  Consider 
\eqn\hom{
\to {\rm Hom}(G, V)\to {\rm Hom}(G, W)\to {\rm Ext}^1(G,U)}
showing that the obstruction to lifting an element of ${\rm Hom}(G, W)$
to an element of ${\rm Hom}(G, V)$ lies in ${\rm Ext}^1(G, U)$.
We have a commutative diagram 
\eqn\homs{\matrix{
{\rm Hom}(W, W) &\buildrel{\partial}\over{\to}& {\rm Ext}^1(W, U) \cr
\da & & \da \cr
{\rm Hom}(G, W)& \to& {\rm Ext}^1(G, U)}}
with $\partial(1)=\xi$ the extensions class. So
we conclude a non-zero element of ${\rm Hom}(G, W)$ can be lifted 
to an element of ${\rm Hom}(G, V)$ exactly when the extension class 
$\xi$ is in the kernel of 
\eqn\fmaptwo{f\colon {\rm Ext}^1(W, U)\to {\rm Ext}^1(G, U)}
Thus if $f$ is injective $f(\xi)\neq 0$ and such a lifting does not exist. 

Before specifying the {\it f-map} in our situation let us 
determine $c_1(G_2)$ and show that
$\mu(G_2)=0$ if and only if $c_1(G_2)=0$. From $G_2\rightarrow \pi^*E_2$, 
we have  $(\Lambda^2G_2)^{**}\rightarrow \Lambda^2 \pi^*E_2^{**}$ 
now $rk(\Lambda^2 G_2)=1$ and $(\Lambda^2 G_2)^{**}$ is a 
reflexive torsion free sheaf of rank one (i.e., a line bundle);
furthermore $c_1(G_2)=c_1((\Lambda^2 G_2)^{**})$. 
As $(\Lambda^2 \pi^*E_2)^{**}=:L$ with $c_1(L)=0$ we find
\eqn\cg{c_1(G_2)=-D_2}
with $D_2$ an effective divisor (set $D_2=A_2\si_1+A_2'\si_2-\pi^*\la_2$). 
The slope of $G_2$ is then given by \slopc\ 
and as $\la_2 c_1\leq 0$ and $A_2+A_2'\geq 0$ we get $\mu(G_2)=0$ 
if and only if $c_1(G_2)=0$. 
 
Let us specify the {\it f-map} in our case. Consider the exact sequence
\eqn\sec{0\to G_2(D)\to \pi^*E_2(D)\to T(D)\to 0}
where $T=\pi^*E_2/G_2$ is a torsion sheaf. 
Applying $Hom(\cdot, \pi^*E_1\ox\cO_X(-2D))$ we get
\eqn\assec{
Ext^1(T(D),  \pi^*E_1(-2D))\to Ext^1(\pi^*E_2(D),\pi^*E_1(-D))
\buildrel{f}\over\to Ext^1(G_2(D),\pi^*E_1(-D))}
Thus we have to show that $Ext^1(T(D),  \pi^*E_1(-D))=0$. We have
\eqn\exts{\eqalign{
Ext^1(T(D),  \pi^*E_1(-D))&=Ext^2( \pi^*E_1(-D),T(D))^*
=H^2(X, \pi^*E_1^*(2D)\ox T)^*}}
We know $\mu(G_2)= 0$ if and only if $c_1(G_2)=0$.
As $T$ is a torsion sheaf we have $T=j_*T'$ 
with $j\colon Y\to X$ and $T'$ some sheaf on $Y$ 
and the Grothendieck-Riemann-Roch theorem gives 
$0=c_1(T)=c_1(j_*T')=j_*(rk(T'))=rk(T')Y_{co\, 1}$.
Thus $T$ is supported in codimension $\geq 2$ and $H^2$ vanishes. 
So destabilizing subsheaves of type $(0,2)$ with $\mu(G_2)=0$ do not occur. 

\subsec{The Bogomolov Inequality}

As a check on our proof of stability we will derive that 
the Bogomolov inequality 
\eqn\bogoin{c_2(V_4)J\geq 0} 
for a stable bundle 
with $c_1(V_4)=0$ is fulfilled. This becomes in our case 
\eqn\bogo{\eqalign{{1\over 2}c_2(V_4)J&={1\over 2}\Big( (k_1+k_2)F
-2\Big[x(2\al -xc_1)\Si +\al^2\Big]\Big)(z\Si + hc_1)\cr
&= -(h-z)2x\Big((\al-xc_1)c_1+\al c_1\Big)-2z\al^2+z(k_1+k_2)\; \geq \; 0}}
We have $x>0$, $h-z>0$, $\al c_1\leq 0$ and $\al-xc_1<0$ showing 
that $c_2(V_4)J\geq 0$ {\it if one is in the case $\al^2\leq 0$}. 
In the general case
\uppb\ becomes (with the notation $\beta:= -\al$)
\eqn\condchain{\eqalign{h^2-(h-z)^2\; <\; 
{xc_1^2 \over xc_1^2 +\beta c_1}\; h^2 
\;\;\;\; \Longleftrightarrow \;\;\;\; 
h^2\; \beta c_1 \; < \; (h-z)^2 (xc_1^2 +\beta c_1)\cr
\Longleftrightarrow \;\;\;\; {h \over h-z} \; < \; 
\Big( {h \over h-z} \Big)^2 \; < \; {xc_1^2+\beta c_1 \over \beta c_1}
\leq {x^2c_1^2+2x\beta c_1 + \beta ^2 \over \beta^2}}}
using the inequality (note that $c_1$ is here ample)
\eqn\ineq{\beta^2 c_1^2 \leq (\beta c_1)^2} 
and $\beta c_1 \geq 0$
(actually we can even assume $\beta c_1 >0$ as otherwise
$\al^2\leq 0$ by the Hodge index theorem when \bogo\ was clear).
So one has indeed (note $k_i\geq 0$)
\eqn\concl{h \beta^2 \; < \; (h-z)(x^2c_1^2+2x\beta c_1 + \beta ^2)
\;\;\;\;  \Longleftrightarrow \;\;\;\; 
0\; < \; (h-z)x(xc_1^2+2\beta c_1 )-z \beta ^2 }

\newsec{Physical Constraints}

\subsec{Breaking the $SU(5)$ GUT group to the Standard Model Gauge Group}

\noindent
On $X'$ one turns on a ${\bf Z_2}$ Wilson line
of generator ${\bf  1_3} \oplus {\bf - 1_2}$ breaking $H=SU(5)$ to $H_{SM}$
\eqn\wilson{SU(5) \lra H_{SM}=SU(3)_c \, \x \, SU(2)_{ew} \, \x \, U(1)_Y}
(up to a ${\bf Z_6}$).
This gives, from ${\bf \bar{5}}=\bar{d}\oplus L$ and ${\bf 10}=Q\oplus 
\bar{u}\oplus \bar{e}$, the fermionic matter content 
\eqn\StMod{\eqalign{
SM\; fermions&=Q\oplus L \oplus \bar{u}\oplus\bar{d} \oplus
\bar{e}\cr
&=({\bf 3},{\bf 2})_{1/3}\oplus ({\bf 1},{\bf 2})_{-1}\oplus ({\bf
\bar{3}},{\bf 1})_{-4/3} \oplus ({\bf \bar{3}},{\bf 1})_{2/3}  \oplus ({\bf
1},{\bf 1})_2}}
of the Standard model.
$ad_{E_8}$ decomposes under
$G\x H= SU(5)_{str.gr.}\x SU(5)_{gau.gr.}$
\eqn\decomp{\eqalign{
{\bf 248} &=({\bf 5},{\bf 10})\oplus ({\bf \bar{5}}, {\bf \overline{10}})
\oplus ({\bf 10},{\bf \bar{5}})\oplus ({\bf \overline{10}},{\bf 5})
\oplus ({\bf 24},{\bf 1}) \oplus ({\bf 1},{\bf 24})}}
For $SU(5)$ GUT models with matter ${\bf \bar{5}}\oplus {\bf 10}$ in one family
one needs to consider besides the fundamental $V={\bf 5}$
to get the ${\bf 10}$-matter also the $\Lambda^2 V={\bf 10}$ 
to get the ${\bf \bar{5}}$-matter; 
as the ${\bf 10}$ and the ${\bf \bar{5}}$ come in the same number of
families (as also demanded by anomaly considerations)
it is enough to adjust $\chi(X,V)$ to get all the Standard model fermions.

All the above considerations concern the net generation number, i.e., 
the number of generations minus the number of anti-generations.
Beyond the mentioned multiplets ideally a string model should
provide no further exotic matter multiplets of nonzero net generation number
(conjugate pairs should pair up and become massive at the string scale).
In a model resembling the MSSM one furthermore wants to have just
one conjugate pair $H, \bar{H}$ of Higgs doublets.
Their pairing is described field-theoretically by the $\mu-term$
$\mu H \bar{H}$ where one has to understand that $\mu$ sits at the
electro-weak scale and not at the string scale, say, when coming from
a string model; in that case the coupling is field-dependent and mediated
by a superpotential term $\la \phi H\bar{H}$ where $\la$ is
the coupling constant and $\phi$ a  superfield which is, just like
the right-handed natutrino $\nu_R$, a singlet under the Standard Model 
gauge group, for example a modulus. If the latter aquires a vev 
(there may be an additional superpotential coupling ${\bf 1^3}$ for $\phi$)
it provides an effective $\mu$-term.

\subsec{Building the Rank $5$ Bundle from the Rank $4$ Bundle}

We explain now more precisely 
how to embed the structure  group into $E_8$ and to get an
$SU(5)$ GUT group (cf. also \refs{\timoone,\timotwo}). First we twist with a line bundle $\cO(-\pi^*\beta)$
and build a split extension (direct sum) to embed the resulting
$SU(4)\ox U(1)$ bundle into $E_8$
\eqn\rankf{
0\to V_4\ox \cO_X(-\pi^*\beta)\to V_5\to \cO_X(4\pi^*\beta)\to 0}
The bundle $V_5$ has the Chern classes
\eqn\chrk{\eqalign{
c_2(V_5)&=-2x(2\al-xc_1)\Si-2\al^2-10\beta^2+k_1+k_2\cr
{c_3(V_5)\over 2}&=2x\Big[k_1-k_2-4\al\beta+2x\beta c_1\Big]}}

Note that $V_5$, as a direct sum, can not be stable, but only poly-stable,
(direct sum of stable bundles of the same slope). This common
slope must be zero as, quite generally, 
for a rank $n$ bundle $\cV=\oplus \cV_i$ composed of $U(n_i)$ bundles of
slopes $\mu_i={1\over n_i}\int J^2 c_1(\cV_i)=:\mu$ (they must coincide for
$\cV$ to be polystable)
one finds $\int J^2 c_1(\cV_i)=0$ for all $i$ as
$0=\int J^2 c_1(\cV)=\sum r_i \mu_i=\mu n$.
For us, having $\mu_i=\beta J^2=2z(2h-z)\beta c_1$, 
this means 
\eqn\slope{\beta \; c_1=0}

More generally one gets the condition $\pi^*\beta\cdot J^2=0$. 
Its possible violation
may be understood either as pointing to the necessary
inclusion of one-loop effects so that the DUY condition is fulfilled
quantum-mechanically \refs{\blume}, \refs{\BfibreIII}, \refs{\Radu};
alternatively the resulting instability can be interpreted \refs{\Radu}
as pointing to a dynamical vacuum shift, i.e., indicating that the stable
vacuum to be considered is actually a {\it non-split} extension \rankf.

\subsec{$U(n)$ Bundles and Line Bundle Twists of $SU(n)$ Bundles}

From an $SU(n)$ bundle $V$ and a line bundle $\cO_X(-\pi^*\beta)$ 
one can build a $U(n)$ bundle by the twist
\eqn\twist{
\cV=V\otimes \cO_X(-\pi^*\beta)}
So $c_1(\cV)\equiv 0 \, (n)$ as $\pi^*\beta$ is integral. Conversely,
if $c_1(\cV)\equiv 0 \, (n)$,
one can split off an integral class $\pi^*\beta$ of $c_1(\cV)=n\pi^*\beta$
and define a corresponding line bundle $\cO_X(-\pi^*\beta)$ 
such that $V:=\cV\otimes
\cO_X(\pi^*\beta)$ is an $SU(n)$ bundle,
i.e., one can think of $\cV$ then as $V\ox\cO_X(-\pi^*\beta)$.

Note that the structure group $U(n)$ arises in the decomposing case 
from $SU(n)\cdot U(1)$ 
(the latter factor is understood here always as embedded by multiples
of the identity matrix)
whereas for a bundle $V\oplus \cL(D)$ the structure group would be the
direct product $SU(n)\times U(1)$. Note that there is a morphism  $f\colon
SU(n)\times U(1) \to U(n)$
sending  $(a,b)\mapsto a\cdot b$. The image of this morphism is $U(n)
=SU(n) \cdot U(1)$, so $SU(n)\cdot U(1) = \Big( SU(n) \times U(1)
\Big)/{\rm ker}(f)$. The subgroup ${\rm ker}(f)$ is formed by all
pairs $(\lambda \cdot {\rm Id_n}, \lambda^{-1})$ where $\lambda \in
{\bf C}$ with $\lambda \cdot {\rm Id_n} \in SU(n)$, i.e.,
$\lambda^n=1$ and ${\rm ker}(f) = {\bf Z_n}$ (the group of $n$-th roots of
unity). As the difference between the direct
product and the product is just a discrete group,
and since all group theoretical statements in this
paper are understood on the level of Lie-algebras, we will write
$SU(4)\times U(1)$ instead of $SU(4)\cdot U(1)$ for our structure group $G$.

\subsec{$E_8$ Embedding and Massive $U(1)$}

Let us make the embedding of $G$ in $E_8$ more explicit.
One embeds a $U(4)$ bundle block-diagonally via
\eqn\embE{U(4)\ni A \to {\left(\matrix{ A&0\cr 0&\det^{-1}A
}\right)}
\in SU(5)}
Therefore, after making the twist 
$\cV=V_4\ox \cO_X(-\pi^*\beta)$ with $c_1(\cV)=-4\pi^*\beta$
one actually has to work with the bundle 
\eqn\Vfive{V_5\; =\; \cV\oplus \cO_X(4\pi^*\beta)\; 
=\; V_4\ox \cO_X(-\pi^*\beta)\; \oplus \; \cO_X(4\pi^*\beta)}
The unbroken gauge group will then be given by $H=SU(5)\times U(1)_X$, 
the commutator of $G=SU(4)\times U(1)_X$ in the observable $E_8$.
The decomposition 
$ad(E_8)=\bigoplus_i U_i^{SU(4)}\otimes R_i^{SO(10)}=
\bigoplus_i (U_i, R_i)=\bigoplus_i (U_i, S_i^{SU(5)})_{t_i^{U(1)}}$
of 
the adjoint representation of $E_8$ under $SU(4)\times SU(5)\times U(1)_X$ 
specifies itself as follows
\eqn\breaks{\eqalign{
{\bf 248} &\longrightarrow
({\bf 5},{\bf 10})\oplus 
(\overline{\bf 5},\overline{{\bf 10}})\oplus ({\bf 10}, \overline{{\bf
5}})\oplus (\overline{\bf 10}, {\bf 5})\oplus ({\bf 24},{\bf 1}) \oplus ({\bf
1},{\bf 24})\cr
 &\longrightarrow
\Big(({\bf 4},{\bf 1})_{-5} \oplus ({\bf 4}, \overline{\bf 5})_3 \oplus
({\bf 4},{\bf 10})_{-1}\Big)
\oplus\Big((\overline{\bf 4},{\bf 1})_{5} \oplus 
(\overline{\bf 4}, {\bf 5})_{-3}
\oplus (\overline{\bf 4},\overline{\bf 10})_{1}\Big)\cr
&\ \ \ \ \ \ \oplus ({\bf 6},{\bf 5})_2 \oplus ({\bf 6},\overline{\bf 5})_{-2}
\oplus ({\bf 15},{\bf 1})_0 \oplus ({\bf 1}, {\bf 1})_0 \oplus 
({\bf 1},{\bf 10})_4 \oplus
({\bf 1}, \overline{\bf 10})_{-4} \oplus ({\bf 1}, {\bf 24})_0
}}
The $SU(5)$ representations are given as an auxiliary step.
The full deomposition, identical to an auxiliary $SU(4)\times SO(10)$ step,
leads to the right-handed neutrino $\nu_R$;
additionally (besides the gauge bosons 
$({\bf 1}, {\bf 1})_0 \oplus ({\bf 1}, {\bf 24})_0$ of $H$)
some neutral matter given by singlets (moduli) arises from $End(V)$, i.e.,
$({\bf 15},{\bf 1})_0$.
The massless (charged) matter content is 
\eqn\mattercontent{\bigoplus (S_k)_{t_k} ={\bf
1_{-5}} \oplus {\bf \bar{5}_3} \oplus {\bf 10_{-1}}\oplus {\bf
\bar{5}_{-2}}\oplus {\bf 10_4}} 
The first three multiplets refer to the $\nu_R$ plus the SM fields.
The ${\bf \bar{5}_{-2}}$ refers to Higgses
related to $\Lambda^2 V$; the last multiplet ${\bf 10_4}$ describes further
exotic matter. Ideally one would want, of course, to avoid net generations
of the last two multiplets.

Precisely those $U(1)$'s in the gauge group $H$ 
which occur already in the structure group $G$ (so-called
$U(1)$'s of type I, other $U(1)$'s in $H$ are called to be of type II)
are anomalous \refs{\blume-\Distler\Dine\Lukas}.
The anomalous $U(1)_X$ can gain a mass by absorbing some of the
would be massless axions via the Green-Schwarz mechanism, that is,
the gauge field is eliminated from the low energy spectrum by
combining with an axion and so becoming massive. One has to check
that the anomalies related to $U(1)_X$ do not cancel accidentally
(i.e., that the mixed abelian-gravitational, the mixed 
abelian-non-abelian and the pure cubic abelian anomaly do not all vanish).
Computing the anomaly-coefficients of $U(1)_X$ one finds 
for \Vfive\
\eqn\mass{\eqalign{
A_{U(1)-G_{\mu\nu}^2}&=\sum tr_{(S_k)_{t_k}}q \, \cdot \chi(X, U_k \ox t_k)
=10 \beta \cdot \Big( 12c_2(V) - 5c_2(X)\Big)\cr
A_{U(1)-SU(5)^2}&=\sum q_{t_k} C_2(S_k)\, \cdot \chi(X, U_k \ox t_k)
=10 \beta \cdot \Big(2c_2(V)- c_2(X)\Big)\cr
A_{U(1)^3}&=\sum tr_{(S_k)_{t_k}}q^3 \, \cdot \chi(X, U_k \ox t_k)
=600 \beta \cdot \Big( 2c_2(V)- c_2(X)\Big)}}
(with $C_2$ normalised to give $C_2(\bar{f})=1$, $C_2(\Lambda^2 f)=3$ for
$SU(5)$).
The first condition is generically independent, 
so not all three coefficients would vanish.
Note that in our specific case,
after taking into account the condition $\beta c_1=0$, one has 
with 
\eqn\evaluations{\beta c_2(V)\; = \; 
\beta \cdot (-4x) (2\al - x c_1) = -8x\al \beta\;\;\; , \;\;\;\;\;\;\;\;\;
\beta c_2(X)\; = \; 0}
that all three expressions will be proportional to
$\al \beta$. So we get for the decoupling of the additional
$U(1)$ in the gauge group the condition
\eqn\decopulingcond{\al\, \beta \neq 0}

\subsec{Avoiding Exotic Matter}

Recall the massless (charged) matter content 
\eqn\matter{\bigoplus (S_k)_{t_k} 
={\bf 1_{-5}} \oplus {\bf \bar{5}_3} \oplus {\bf 10_{-1}}\oplus {\bf
\bar{5}_{-2}}\oplus {\bf 10_4}} 
Here we have a Standard Model fermion generation 
${\bf \bar{5}_3} \oplus {\bf 10_{-1}}$ 
plus a right-handed neutrino ${\bf 1_{-5}}$.
Besides these multiplets
the ${\bf \bar{5}_{-2}}$ refers to Higgses related to $\Lambda^2 V$; 
furthermore we have exotic matter ${\bf 10_4}$ which we would like to avoid.
The net-amount of such states (which could not pair up and become massive)
is computed from $\chi(X, U_k\ox t_k)$
\vskip .5cm

\centerline{
\vbox{\hsize=2truein\offinterlineskip\halign{\tabskip=2em plus2em
minus2em\hfil#\hfil&\vrule#&\hfil#\hfil&\vrule#&\hfil#\hfil\tabskip=0pt\cr 
matter multiplet  &depth6pt& net-amount\cr 
\noalign{\hrule\hrule}
$({\bf 4},{\bf 10})_{-1}$ &height14pt& $\chi(X, V\ox \cL^{-1})$\cr
$({\bf 1},{\bf 10})_4$  &height14pt& $\chi(X, \cL^{4})$\cr
$({\bf 4}, \bar{\bf 5})_3$  &height14pt& $\chi(X, V\ox \cL^{3})$\cr
$({\bf 6},\bar{\bf 5})_{-2}$ &height14pt& $\chi(X, \Lambda^2 V\ox \cL^{-2})$\cr
$({\bf 4},{\bf 1})_{-5}$  &height14pt& $\chi(X, V\ox \cL^{-5})$\cr
\noalign{\medskip} }}} \nobreak

One gets for the individual terms with $\beta c_1=0$
\eqn\individ{\eqalign{
\chi(V\ox \cL^{-1}) &={c_3(V)\over 2}-\beta \Big({c_2(X)\over 3}-c_2(V)\Big)
=2x\Big(k_1-k_2-4\al \beta\Big)\cr
\chi(\cL^4)&=\;\;\;\;\;\;\;\;\;\;\;\;\;\; \beta {c_2(X)\over 3}
\;\;\;\;\;\;\;\;\;\;\;\;\;\;\;\;\;\; =0\cr
\chi(V\ox \cL^{3}) &={c_3(V)\over 2}+\beta \Big(c_2(X)-3 c_2(V)\Big)
= 2x\Big(k_1-k_2+12\al \beta\Big)\cr
\chi(\Lambda^2V\ox \cL^{-2})&=\;\;\;\;\;\;\;\;\;\;\;
-\beta \Big(c_2(X)-4 c_2(V)\Big)
=2x\Big( -16 \al\beta\Big)\cr
\chi(V\ox \cL^{-5})&={c_3(V)\over 2}-5\beta\Big( {c_2(X)\over 3}-c_2(V)\Big)
=2x\Big(k_1-k_2-20\al \beta\Big)
}}
Note as a check that the non-abelian anomaly vanishes
\eqn\check{
\chi(V\ox \cL^{-1})+\chi(\cL^{4})
-\chi(V\ox \cL^{3})-\chi(\Lambda^2V \ox \cL^{-2})
= 0}
Equivalently the net-chirality 
$\chi(V\ox \cL^{3})+\chi(\Lambda^2V\ox \cL^{-2})$
of the ${\bf \bar{5}}$-matter
and $\chi(V\ox \cL^{-1})+\chi(\cL^{4})$
of the ${\bf 10}$-matter coincide as they should.
Avoiding a non-zero net chirality from the Higgs sector
(also there would be no mass terms for unpaired ${\bf 10}_4$'s)
we may demand 
\eqn\avoid{
N_{gen}(\Lambda^2V \ox \cL^{-2})=0=N_{gen}(\cL^{4})}
Interpreting these generation numbers 
as indices gives the vanishing conditions
$N_{gen}(\cL^{4})=0$ (which means $\beta c_2(X)=0$ and 
is automatically fulfilled for $\beta c_1=0$) and furthermore
$N_{gen}(\Lambda^2 V \ox \cL^{-2})=-32x\al \beta =0$
(which in view of the previous condition means $\beta c_2(V)=0$).
Furthermore one has then just 
$N_{gen}= \chi(V\ox \cL^{-1})=  c_3(V)/ 2$.

One gets from $N_{gen}(\Lambda^2 V \ox \cL^{-2})=0$ the condition
\eqn\avoidexoticmatter{\al \beta \; =\; 0}
Therefore one has to face the following alternative: either one 
proceeds as described in \avoidexoticmatter\ and avoids the exotic matter,
then the $U(1)$ in the gauge group has not decoupled; or, 
alternatively, one insists on this decoupling, i.e. on the condition
\decopulingcond\ that the anomaly is present, but keeps then
 the exotic matter.
(An extended analysis shows that the situation is not improved by considering 
$D'=y\Si + \pi^*\beta$ with $y \neq 0$.)

\newsec{List of Constraints and Solutions}

Let us list all the constraints we have found. 
Keep in mind that we assume always $x>0$ and $k_1, k_2\geq 0$; 
furthermore we implement \slope\ (note also \chrk)
\eqn\conditions{\eqalign{
\al - x c_1 \; & < \; 0\cr
0\; &<\; {1\over 2}\chi=2+2c_1^2-6\al c_1+4\al^2-(k_1+k_2) 
\ \ \ \ \ \ {\rm for}\ \ \ \ x=1\cr
\al c_1\; &\leq \; 0 \cr
w_B&=(6-2x^2)c_1+4x\al\geq 0 \;\;\;\; \;\;\;\; 
\buildrel{x=1}\over \longrightarrow  \;\;  \al \geq -c_1 \cr
a_f&=44+2\al^2+10\beta^2-(k_1+k_2)\geq 0\cr
{1\over 2}N_{gen}&=x\Big(k_1-k_2-4\al\beta\Big)=\pm 3}}
From $w_B\geq 0$ together with $\al c_1 \leq 0$ 
(such that $\al =(p,q)$ can not have $p,q>0$) one finds 
\eqn\xfix{x=1}
In writing $N_{gen}$ we have already used the slope condition $\beta c_1=0$ 
from \slope.
Therefore, on ${\bf F_0}={\bf P^1}\times {\bf P^1}$ 
one gets explicitely that $\beta = (e, -e)$, or 
more precisely with $a_f\geq 0$ 
\eqn\bet{\beta \; = \; \pm (1, -1)}
(using that $\al^2\leq 4$).
Furthermore, one has the following reflection property
\eqn\reflection{
\Big( \al = (p,q), \; (k_1, k_2)\longrightarrow N_{gen}=\pm 6\Big)
\Longrightarrow
\Big( \al = (q,p), \; (k_2, k_1)\longrightarrow N_{gen}=\mp 6\Big)}

One has to distinguish two cases. One can implement either the condition
$\al \beta =0$ avoiding so exotic matter 
but keeping the additional $U(1)$ in the low-energy spectrum.
This is the case we are going to describe now.
Alternatively, one can choose the inverted option
(keeping exotic matter but avoiding the additional $U(1)$)
if $\al \beta \neq 0$; this is discussed further in the appendix $D$.

So let us study here the case $\al \beta =0$ where we avoid the exotic matter
(but keep an additional $U(1)$).
By  the conditions given above this means $\al=(p,p)$ with 
$p=-2, -1$ or $0$. The latter case does not actually occur
when taking into account the bounds $k_i \geq 8$:
for $\al =(0,0)$ one would get 
$k_1+k_2<18$ from the non-split condition $\chi>0$
which is not solvable together with $k_1-k_2=\pm 3$ and $k_i\geq 8$.
The latter condition is forced on us (for $h=1/2$) as we want to use the 
concrete bundles $E_i$ on the base $B$, constructed in appendix $B$.

The reflection property \reflection\ implies that 
for each solution pair $(k_1, k_2)$ also the further pair $(k_2, k_1)$
occurs (by which the following list has to be augmented).
\bigskip\centerline{
\vbox{\hsize=2truein\offinterlineskip\halign{\tabskip=2em plus2em
minus2em\hfil#\hfil&\vrule#&\hfil#\hfil&\vrule#&\hfil#\hfil
&\vrule#&\hfil#\hfil\tabskip=0pt\cr 
$\al$ &depth6pt& $k_1$ &depth6pt& $k_2$ &depth6pt& $i$  \cr 
\noalign{\hrule}
$(-2,-2)$ &height14pt&  $8+i$ &height14pt&  $11+i$
&height14pt& $i=0, \dots , 10$  \cr
$(-1,-1)$ &height14pt&  $8+i$ &height 14pt&  $11+i$
&height14pt& $i=0, \dots , 4$  \cr
\noalign{\medskip} }}} \nobreak
Here the relative size $k_1-k_2$ of the instanton numbers is given by 
the physical generation number $N_{gen}^{phys}=N_{gen}/2$
(downstairs on $X'=X/{\bf Z_2}$). The lower bound on $i$ comes from 
the conrete construction of the $E_i$ (cf.~appendix)
and the upper bound stems from the condition 
$a_f\geq 0$ (which is stronger than the non-split condition $\chi >0$), 
$k_1+k_2\leq 24 + 2\al^2$.

\newsec{Conclusions}

We build heterotic sting models with the gauge group
of the Standard Model times an additional $U(1)$. The net amount of chiral
matter is given by precisely 
three net generations of chiral fermions of the Standard Model
(including a right-handed neutrino  $\nu_R$).
This is done by building first 
the elliptic Calabi-Yau space $X$ over ${\bf F_0}={\bf P^1}
\times {\bf P^1}$, the 
$B$-fibration with two sections $\si_1$ and $\si_2$ 
possessing a free involution $\tau_X$ leaving the holorphic three-form 
invariant.
Then we construct 
an invariant $SU(5)$ model of six net generations over $X$ which descends 
to the quotient Calabi-Yau $X'=X/{\bf Z_2}$ and is afterwards broken 
to the Standard Model by turning on a Wilson line;
the latter is possible as $\pi_1(X')={\bf Z_2}$.

The invariant $SU(5)$ model on the cover space $X$ arises 
actually not from an invariant $SU(5)$ bundle but from
an $SU(4)\times U(1)$ bundle (where $\beta = \pm (1,-1)$)
$$V_5=V_4\otimes \cO_X(-\beta)\oplus \cO_X(4\beta)$$

The invariant rank 4 bundle $V_4$ of $c_1(V_4)=0$ 
is defined by a non-split extension
$$0\to \pi^*E_1\ox \cO_X(-D)\to V\to \pi^*E_2\ox \cO_X(D)\to 0$$
Here $D$ is the divisor $D=\Si +\pi^*\al$, chosen invariant, where $\al$
is one of the elements $(-2,-2)$ or $(-1,-1)$,
and the $E_i$ are two stable vector bundles on the base $B$ 
of rank two, $c_1(E_i)=0$ and $c_2(E_i)=k_i$; 
concretely they can be described by the construction of the appendix
for $k_i\geq 8$.
The list of applicable instanton numbers
$(k_1, k_2)$ is given in table 1.
The physical generation number $N_{gen}^{phys}=N_{gen}/2$
(downstairs on $X'=X/{\bf Z_2}$) 
is given just by the mismatch of the instanton numbers
\eqn\mainform{N_{gen}^{phys}=k_1-k_2}
These data-sets fulfil all necessary conditions for the existence
of a non-split extension, stability, DUY-equation,
fivebrane effectivity and generation number.

The information about the matter content 
concern here just the fermions, and further only their net generation number.
Extensions of these investigations describing cases with the appropriate
Higgs content and the individual numbers of generations and anti-generations
will be reported elsewhere.

\bigbreak\bigskip
B. A. thanks H. Kurke and D. Hern\'andez Ruip\'erez for discussions.

\appendix{A}{Cohomological Relations}

\bigskip\noindent{\it Relations from the Leray Spectral Sequence}

The Leray spectral sequence of the fibration $\pi:X\rightarrow B$
relates the cohomology of a bundle $V$ on $X$ to the cohomology 
of the higher direct image sheaves $R^i\pi_*V$ on the base $B$. 
The latter are defined by 
$R^i\pi_*V({\cal U}) = H^1(\pi^{-1}({\cal U}), V\vert_{\pi^{-1}({\cal U})})$
for an open set ${\cal U}\subset B$; moreover, for any point $b\in B$ 
one has $R^i\pi_*V\vert_b=H^i(F_b, V\vert_{F_b})$. 
Applied to our situation in section 4.2 the Leray spectral sequence 
degenerates to the long exact sequence
\eqn\leray{\eqalign{
0&\to H^1\Big(B, \cE\ox \cO_B(-2\al)\ox\pi_*\cO_X(-2x\Si)\Big)\to 
H^1\Big(X, \pi^*\cE \ox \cO_X(-2D)\Big)\to\cr
&\to H^0\Big(B, \cE\ox \cO_B(-2\al)\ox R^1\pi_*\cO_X(-2x\Si)\Big)\to \cr
&\to H^2\Big(B, \cE \ox \cO_B(-2\al)\ox\pi_*\cO_X(-2x\Si)\Big)
 \to H^2\Big(X, \pi^*\cE \ox \cO_X(-2D)\Big)\to\cr
 &\to H^1\Big(B, \cE \ox \cO_B(-2\al)\ox R^1\pi_*\cO_X(-2x\Si)\Big)\to 0}}
(using the projection formula $R^i\pi_*(V\ox \pi^*M)=R^i\pi_*(V)\ox M$)
together with 
\eqn\leraytwo{\eqalign{H^0\Big(X, \pi^*\cE \ox \cO_X(-2D)\Big)&=
H^0\Big(B, \cE \ox \cO_B(-2\al)\ox \pi_*\cO_X(-2x\Si)\Big)\cr
H^3\Big(X, \pi^*\cE \ox \cO_X(-2D)\Big)&=H^2\Big(B, \cE
\ox \cO_B(-2\al)\ox R^1\pi_*\cO_X(-2x\Si)\Big)}}

\bigskip\noindent{\it Splitting Relations}

Let us prove the following relations
\eqn\relsp{\eqalign{\pi_*\cO_X(y\Si)&=\cO_B\oplus K_B\oplus2K_B^2
\oplus\dots\oplus 2K_B^y, \ \ y>1\cr
R^1\pi_*\cO_X(-y\Si)&=\cO_B\oplus K_B\oplus2K_B^{-1}\oplus\dots
\oplus 2K_B^{1-y}, \ \ y>1}}
We start from the exact sequence
\eqn\stepo{0\to \cO_X(\si_1)\to \cO_X(\Si)\to \pi^*K_B\vert_{\si_2}\to 0}
if we apply $R^i\pi_*$ and recall $R^1\pi_*\cO_X(\si_1)=0$ 
as well as $\pi_*\cO_X(\si_1)=\cO_B$ we find 
\eqn\stepto{0\to\cO_B\to \pi_*\cO_X(\Si)\to K_B\to 0}
which splits (as $Ext^1(K_B, \cO_B)=H^1(B, K_B^*)=0$) and so we get
\eqn\reso{\pi_*\cO_X(\Si)=\cO_B\oplus K_B}
The rest of the above formula follows by induction using ($y>1$)
\eqn\steth{0\to \pi_*\cO_X((y-1)\Si)\to 
\pi_*\cO_X(y\Si)\to\pi_*\cO_X(y\Si)\vert_\Si\to 0}
and that $\pi_*\cO_X(y\Si)\vert_\Si=2K_B^y$ 
(note that $\si_1$ and $\si_2$ are disjoint). 
The relation for $R^1\pi_*\cO_X(-y\Si)$ 
follows from relative duality and using the fact 
that the right hand side is locally free such that we can take the 
double dual of the left hand side\foot{note that
since for every point  $p\in B$ we have  $h^1(F_p, \cO_{F_p}(-\si_1(p)))=1$, 
the function   $p\mapsto h^1(F_p, \cO_{F_p}(-\si_1(p)))$ is constant 
and then $R^1\pi_*\cO_X(-\si_1)$ is locally free of rank 1 (cf. \refs{\Hart}, 
Cor. III.12.9); the same reasoning applies for $R^1\pi_*\cO_X(-y\Si)$}
\eqn\serr{[R^1\pi_*\cO_X(-y\Si)]^{*}=\pi_*\cO_X(y\Si)\ox K_B^{*}}
\noindent
{\it Remark:}
Let us close with the following remark concerning the case with $x> 0$. One has
\eqn\negx{x>0\, : \;\;\;\;\;\;\;\;
\pi_*\cO_X(-y\Si)=0 \, , \;\;\;\; R^1\pi_*\cO_X(-y\Si)\neq 0}
as $\pi_*\cO_X(-y\Si)\vert_b=H^0(F_b, \cO_{F_b}(-2y[p]))=0$ 
since $-2y[p]$ is the negative of a non-zero effective divisor;
as the degree of that divisor is negative one has
$H^1(F_b, \cO_{F_b}(-2y[p]))\neq 0$ by the Riemann-Roch theorem 
(or by $ H^1(F_b, \cO_{F_b}(-2y[p]))=H^0(F_b, \cO_{F_b}(2y[p]))\neq 0$
as $2y[p]$ is a non-zero effective divisor).

\appendix{B}{Stable Bundles on the Base}

To construct explicitly an appropriate rank $r$ vector bundle $V$ on $B$ 
we consider a suitable twisted extension of $\oplus_{i=1}^{r-1} I_{Z_i}$
by $\cO_B$. Here the $I_{Z_i}$ are ideal sheaves of point sets $Z_i$.
The bundle $V$ should be stable with respect to a certain ample divisor $H$ and
$H$ enters already the construction in form of the twist bundle.
So consider the extension
\eqn\modser{0\; \to \; \cO_B(-(r-1)H)\; \to \; V \; \to \;
\cO_B(H)\ox \; \bigoplus_{i=1}^{r-1} I_{Z_i} \; \to \; 0}
$V$ is known \refs{\Li} to be an $H$-stable bundle of $c_1(V)=0$ and 
\eqn\cherclser{c_2(V)=l(Z)-{r(r-1)\over 2} H^2}
if $l(Z)=\sum_{i=1}^{r-1}l(Z_i)$ fulfills the bound below. 
For $i=1,\dots, (r-1)$ one can choose a reduced 
$0$-cycle $Z_i=Z_i'\cup Z_i''$ with $Z'_i$  
is a generic $0$-cycle in the Hilbert scheme $Hilb^{l(Z_i')}(B)$ 
(so $l(Z'_i)$ has to be sufficiently large, i.e., 
$l(Z'_i)\geq {\rm max}(p_g, h^0(B, \cO_B(rH+K_B)))$; this also guarantees 
that $Z'_i$ satisfies the Cayley-Bacharach property with respect to 
$|rH+K_B|$). Moreover, $Z''_i$ is a  
reduced $0$-cycle (chosen suitably generic)
of length $l(Z''_i)\geq 4(r-1)^2H^2$. Then one has
for a surface with $p_g(B)=0$ (like ${\bf F_0=P^1\x P^1}$) 
\eqn\bounded{l(Z)=\sum_{i=1}^{r-1} l(Z_i')+l(Z''_i)
\geq (r-1)\Big(1+h^0(B, \cO_B(rH+K_B))+4(r-1)^2H^2\Big)}
Applied to our situation where $H=hc_1$ and $r=2$ one gets
(for $h=1/2$)
\eqn\boundtt{\eqalign{c_2(E_i)&=k_i=l_2(Z)-2\cr
l_2(Z)&\geq 2+8(2h-1)h+32h^2=10}}

\bigskip\noindent{\it Invariance of $E_i$}

The involution on the base is just $\tau_B: (z_1, z_2)\lra (-z_1, -z_2)$
(written in affine coordinates).
The input data in the defining extension of $E_i$ are just divisors and
point sets. Therefore it suffices to show that one can first
choose these objects themselves invariant,
and then to assure the existence of an invariant extension.

For divisors like $D=\cO_{{\bf F}_0}(pb+qf)$ one just selects 
for the defining equation those monomials $z_1^rz_2^s$ with $r+s$ even. 
For the point sets of $l(Z)=2k$ elements one chooses
$k$ 'mirror-pairs' of points $\{x, \tau_B(x)\}$
(for an odd number of points one adjoins a fixed point).

In the explicit construction a genericity requirement on the points arises 
besides the bound on the cardinality shown in \boundtt.
Furthermore, as just described, 
the invariance of the bundles will need among other requirements
the invariance of the point set; i.e., the point set has to consist
of mirror pairs (plus a fixed point if the cardinality is odd; the four
fixed points are not generic, so we prefer to work with an even
cardinality).
One convinces oneself that the concrete genericity requirements 
needed in the construction of \refs{\Li} are not violated
by the fact that the point set $Z$ consists of mirror pairs as dictated by
the invariance requirement; for the required sort of genericity
in these arguments is of the type 'choose from an open set' (say a 
complement of divisors) and does not restrict the mutual positions 
of the points).

For the extension note that the action of the involution $\tau_B$
breaks the space of extensions into invariant and anti-invariant parts. 
To guarantee the invariance of an extension we need  
to check that the dimension of the $+$ eigenspace is non-vanishing.
For this we need to describe the appropriate action on $Ext^1$.
Note that the $\tau$ action on the extension bundle 
is reflected by the $\tau$ action on $H^1\big(X, Hom(W, U)\big)$.
For invariant bundles $W, U$ one can still twist the $\tau$ action
on one of them by the full reflection $v\to -v$ in the fiber vector space,
switching the action on $Ext^1$ by a sign. Therefore as soon as $Ext^1\neq 0$
one can conclude that even an invariant extension bundle exists.

\appendix{C}{The Fixed Point Formula on the Moduli Space}

Let $\cM^H(r, 0, k)$ be the moduli space of bundles $E$ over the surface $B$
(stable w.r.t. $H$) of rank $r$, $c_1(E)=0$ and $c_2(E)=k$.
According to \refs{\Artamk} this space is non-empty if $c_2(E_i)>rk(E_i)+1$
(provided that $p_g(B)=0$ as in our case of $B={\bf F_0}$).
We want to compute the dimension of the moduli space $\cM_+^H(r, 0, k)$
of stable $\tau_B$-{invariant} bundles. 
Recall that $H^0(B, ad \, E)=0$ ($E$ is 'simple') for $E$ stable 
and $H^2(B, ad \, E)=0$ ($E$ is 'good') at smooth points of the moduli space;
one has then $\dim \cM^H(r, 0, k)=\dim H^1(B, ad \, E)$.
Recall also that on a surface $B$ with {\it effective} anti-canonical divisor
a simple $E$ is already good 
(cf.~Ch.~6, Prop.~17 \refs{\Friedmanbook}); in other words $\cM^H(r, 0, k)$ 
(consisting of stable and therefore simple bundles) is smooth.

Assume that the action of the involution $\tau_B$ can be lifted 
to an action on $E$, so the action of $\tau_B$
lifts also to an action on the endomorphism bundle $End(E)$.
The index of the
$\bar\partial$ operator then generalizes to a character valued index where
for each $g\in G$ one defines
${\rm index}(g)=\sum_{i=0}^2(-1)^{i+1}{\rm Tr}_{H^i(X, End(E))}g$. 
For the linear combination ${1+\tau_B\over 2}$ 
which projects onto the even subspace one finds the
${\rm index}({1+\tau_B\over 2})=
\sum_{i=0}^2(-1)^{i+1}{\rm dim}H^i_+(X, End(E))$.

More precisely, recall that the index of the $\bar\partial$ operator
arises from the complex $(E_i)=({\cal A}^{0,i}(B, E))$.
Then there exists a commutative diagram
\eqn\pullbackbundle{\eqalign{
\; \tau_B^* E \; & \buildrel{\phi_{\tau_B}}\over \lra   \; E\cr
\da \;\; &  \;\;\;\;\;\;\;\;\;\, \da \cr
B \;\;& \buildrel{\tau_B}\over \lra \; B}}
(Note that, on the other hand, $E$ is called $\tau_B$-invariant
if such a bundle isomorphism exists which covers the identity in the base.)

This gives rise to corresponding maps $\phi^{(i)}_{\tau_B}$
for the spaces in the complex, and then also to maps
$\Phi^{(i)}_{\tau_B}: \Gamma(E_i)\lra \Gamma(E_i)$
via the compositions $\Gamma(E_i) \buildrel{\phi_{\tau_B}}\over \lra 
\Gamma(\tau_B^* E_i) \buildrel{\phi^{(i)}_{\tau_B}}\over \lra \Gamma(E_i)$.
So there are also induced maps in cohomology and we can speak about the
invariant subspaces $H^i_+(B, End E)$.
What concerns  the other side of the fixed point formula,
one has for $b\in $B
linear maps $(\phi^{(i)}_{\tau_B})_b: (E_i)_{\tau_B(b)}\lra (E_i)_b$,
so one can define in particular the contributions 
${\Tr_{(T^{0,i}_b)^*\ox End E}\tau_B \over \det(1-D\tau_B)|_b}$
for the fixed points.

To compute the dimension of the submanifold 
$\cM^H_+(r, 0, k)\hookrightarrow \cM^H(r, 0, k)$, the strategy is 
to consider the corresponding decomposition of the tangent space of 
$\cM^H(r, 0, k)$ at a point representing an invariant bundle;
for the following discussion let us assume that 
such a point in the moduli space exists.
We will compute the 
dimension $\dim H^1_+(B, ad \, E)=\dim \cM^H_+(r, 0, k)$ of the tangent space 
to the invariant bundles, or in other words the dimension of the invariant 
part of the tangent space.
I.e., by the mentioned unobstructedness, one can detect 
a non-trivial dimension of $\cM^H_+(r, 0, k)$ 
(if one can secure its non-emptyness at all)
by looking at invariant first-order deformations. 
Note that we can not apply this type of argument directly on $X$ 
as there can very well be obstructions to first-order deformations.

Although we will have shown $h^1_+>0$,
we still have to make sure that some invariant bundle $E$ exists;
the corresponding statement that (besides dimensional arguments and 
smoothness) the full moduli space $\cM^H(r, 0, k)$
is non-empty was given in \refs{\Artamk}. 
In other words one would  have to augment 
the abstract existence argument given here by 
one concrete construction of an invariant 
bundle. This was given in the previous appendix.

Note that one has from $End\, E=\cO_B \oplus ad\, E$
that $h^0(B, End\, E)\; = \; h^0(B, ad\, E)+1$ 
and $h^i(B, End\, E)\; = \; h^i(B, ad\, E)$ for $i=1,2$. 

So consider the evaluation using the Atiyah-Bott fixed point theorem
\eqn\equivarindex{\eqalign{
\sum_{i=0}^2(-1)^i \dim H^i_+(B, End \, E)
&=\sum_{i=0}^2(-1)^i\Tr_{H^i(B, End \, E)}{1+\tau_B \over 2}\cr
&={1\over 2}\chi(B, End\, E)
+{1\over 2}\sum_{i=0}^2(-1)^i\Tr_{H^i(B, End E)}\tau_B\cr
&={1\over 2}\int_B ch(E)ch(E^*)td(B)
+{1\over 2}\sum_{\tau_B(b)=b}\sum_{i=0}^2(-1)^i
{\Tr_{(T^{0,i}_b)^*\ox End E}\tau_B \over \det(1-D\tau_B)|_b}\cr
&={1\over 2}(r^2-2rk)+ {1\over 2} \; 4\; {4r^2\over 4}}}
(without the inserted $\tau_B$ action this reasoning gives the dimension 
$2kr-(r^2-1)+h^2(B, ad\, E)$ of the tangent space to the moduli space).
Here $\sum (-1)^i \Tr_{H^i}\tau_B$
is evaluated by the fixed point theorem, 
giving a sum over the four fixed points weighted 
by a suitable determinant.

Recall that on $B={\bf F_0}={\bf P^1}\x {\bf P^1}$ the involution $\tau_B$
is given in local affine coordinates by $(z_1, z_2)\lra (-z_1, -z_2)$.
This has four fixed points at 
$(0,0), (0, \infty), (\infty, 0), (\infty, \infty)$.
The differential being given by the diagonal matrix 
diag(-1, -1) one finds for the denominator contributions
$\det(1-D\tau_B)|_b=4$. For what concerns the numerator note that
\eqn\trace{\eqalign{
\sum_{i=0}^2(-1)^i\Tr_{(T^{0,i}_b)^*\ox End E}\tau_B \; &= \; 
\sum_{i=0}^2(-1)^i\Tr_{(T^{0,i}_b)^*}\tau_B \cdot \dim End E +
\dim (T^{0,i}_b)^* \cdot \Tr_{End(E)}\tau_B\cr
&=\Big( 1 - (-2)+1\Big) r^2 +\Big(1-(+2)+1\Big)\Tr_{End(E)}\tau_B }}
(from the local expressions $1, d\bar{z}_i, d\bar{z}_1\wedge d\bar{z}_2$).
Thus one gets for the fixed point correction term given by the double sum 
$4\cdot {4r^2\over 4}=4r^2$.
So we find for the dimension $\dim H^1_+(B, ad \, E)$ of the space of 
even moduli
\eqn\evenmod{\eqalign{h^1(B, ad \, E)-1&= 2rk-r^2\cr
h^1_+(B, ad \, E)-1&={2rk-r^2\over 2}- 2r^2}}

Therefore we get as condition for the existence of an invariant bundle
\eqn\invarcond{c_2(E)\; \geq \; {5\over 2} r -{1\over r}}
whereas the condition for the existence of a general bundle
was just $c_2(E)\geq {1\over 2}r-{1\over 2r}$.

\appendix{D}{Solutions without an additional $U(1)$ 
but with exotic matter}

In the main body of the paper we have described the solutions
to the system of equations \conditions\ in the case $\al \beta =0$;
this amounts to cancelling any exotic matter beyond the Standard Model 
(plus the right handed neutrino).
But one is paying a price for this: the mentioned condition
is also the one giving the anomaly of the unbroken $U(1)$. Only if the 
latter is anomalous it becomes massive, and so decoupling from the 
low-energy spectrum. This decoupling is inhibited if the exotic matter
is cancelled.

Alternatively one can study the case where the additional $U(1)$ 
really becomes anomalous and decouples if one is willing to keep 
the additional exotic matter. So, to describe the corresponding solutions
let us study the case $\al \beta \neq 0$.
The symmetrical cases
$p=q$ and $k_1=k_2$ are independently excluded by \decopulingcond\
and \nge, respectively. Further $p+q\leq 0$ and $p,q\leq 2$ from
$\al c_1\leq 0$ and $\al \leq c_1$, imply,  
together with $p,q\geq -2$ from $\al \geq -c_1$, 
the list 
\eqn\alphalist{\al \; = \; (-2,-1), \; (-2,0), \; (-1,0)
\;\;\;\;\;\;\;\; {\rm or }\;\; (p\leftrightarrow q) -
{\rm reflections \; of \; these}}
Though not a priori excluded, one finds by inspection of the solutions to 
the full system of conditions that the cases 
$(-2, 1), (-2, 2),(-1, 1), (-1, 2)$ do not occur.

The complete set of solutions for the instanton numbers of the 
two plane bundles on the base is for $\beta = (1,-1)$ given by
(for $\beta = (-1, 1)$ one just interchanges the $k_i$)
\bigskip\centerline{
\vbox{\hsize=2truein\offinterlineskip\halign{\tabskip=2em plus2em
minus2em\hfil#\hfil&\vrule#&\hfil#\hfil&\vrule#&\hfil#\hfil
&\vrule#&\hfil#\hfil\tabskip=0pt\cr 
$\al$ &depth6pt& $k_1$ &depth6pt& $k_2$ &depth6pt& $i$  \cr 
\noalign{\hrule}
$(-2,-1)$ &height14pt&  $\;\; 9+i$ &height14pt& $8+i$ 
&height 14pt& $i=0, \dots , 7$ \cr
$(-2,-1)$ &height14pt& $15+i$ &height14pt& $8+i$ 
&height 14pt& $i=0, \dots , 4$ \cr
$(-2, 0)$ &height14pt& $13+i$ &height14pt& $8+i$ 
&height 14pt& $i=0,  1$ \cr
$(-1, 0)$ &height14pt&  $\;\; 9+i$ &height14pt& $8+i$ 
&height 14pt& $i=0, \dots , 3$ \cr
$(-1, 0)$ &height14pt& $15+i$ &height14pt& $8+i$ 
&height 14pt& $i=0$ \cr
\noalign{\medskip} }}} \nobreak

\listrefs

\bye